\documentclass[fleqn,usenatbib]{mnras}

\usepackage[T1]{fontenc}
\usepackage{ae,aecompl}

\usepackage{graphicx}	
\usepackage{amsmath}	
\usepackage{amssymb}	
\usepackage{float}
\usepackage{natbib}
\usepackage{morefloats}
\usepackage{dcolumn}
\usepackage{amsmath}
\usepackage{latexsym}
\usepackage{longtable}
\usepackage{lipsum}
\usepackage{gensymb} 
\usepackage{adjustbox}
\usepackage{lscape}

\title[SNe 2008aq and 2019gaf]{A Study of Two Type IIb Supernovae: SNe 2008aq and 2019gaf}
\author[Mridweeka Singh et al.]{Mridweeka Singh$^{1}$\thanks{E-mail: mridweeka.singh@iiap.res.in, yashasvi04@gmail.com},
Devendra K. Sahu$^{1}$,
Raya Dastidar$^{2,3,17}$,
Rishabh Singh Teja$^{1,4}$
\newauthor
Anjasha Gangopadhyay$^{5,6}$,
G. C. Anupama$^{1}$,
D. Andrew Howell$^{7,8}$,
\newauthor
K. Azalee Bostroem$^{9}$,
Curtis McCully$^{7,8}$,
Jamison Burke$^{7,8}$,
Arti Joshi$^{10}$,
\newauthor
Daichi Hiramatsu$^{7,8,11,12}$,
Hyobin Im$^{13,14}$,
Shubham Srivastav$^{15}$,
Kuntal Misra$^{16}$
\\
$^{1}$Indian Institute of Astrophysics, II Block, Koramangala, Bengaluru 560 034, India \\
$^{2}$Millennium Institute of Astrophysics (MAS), Nuncio Monsenor Sòtero Sanz 100, Providencia, Santiago, Chile\\
$^{3}$Departamento de Ciencias Fisicas, Universidad Andres Bello, Fernandez Concha 700, Las Condes, Santiago, Chile\\
$^{4}$Tsung-Dao Lee Institute, Shanghai Jiao Tong University, No.1 Lisuo Road, Pudong New Area, Shanghai, China\\
$^{5}$Oskar Klein Centre, Department of Astronomy, Stockholm University, AlbaNova, SE-106 91 Stockholm, Sweden \\
$^{6}$Hiroshima Astrophysical Science Centre, Hiroshima University, 1-3-1 Kagamiyama, Higashi-Hiroshima, Hiroshima 739-8526, Japan \\
$^{7}$Las Cumbres Observatory, 6740 Cortona Drive, Suite 102, Goleta, CA 93117-5575, USA\\
$^{8}$Department of Physics, University of California, Santa Barbara, CA 93106-9530, USA\\
$^{9}$DiRAC Institute, Department of Astronomy, University of Washington, Box 351580, U.W., Seattle, WA 98195, USA\\
$^{10}$Institute of Astrophysics, Pontificia Universidad Católica de Chile, Av. Vicuña MacKenna 4860, 7820436, Santiago, Chile\\
$^{11}$ Center for Astrophysics \textbar{} Harvard \& Smithsonian, 60 Garden Street, Cambridge, MA 02138-1516, USA\\
$^{12}$The NSF AI Institute for Artificial Intelligence and Fundamental Interactions\\
$^{13}$Korea Astronomy and Space Science Institute, 776 Daedeokdae-ro, Yuseong-gu, Daejeon 34055, Republic of Korea\\
$^{14}$Korea University of Science and Technology (UST), 217 Gajeong-ro, Yuseong-gu, Daejeon 34113, Republic of Korea\\
$^{15}$Astrophysics Research Centre, School of Mathematics and Physics, Queen’s University Belfast, Belfast BT7 1NN, UK\\
$^{16}$Aryabhatta Research Institute of observational sciencES, Manora Peak, Nainital 263 001, India \\
$^{17}$Istituto Nazionale di Astrofisica, Osservatorio Astronomico di Brera, via E. Bianchi 46, 23807 Merate (LC), Italy.
}

\date{Accepted XXX. Received YYY; in original form ZZZ}

\pubyear{2026}

\begin{document}
\label{firstpage}
\pagerange{\pageref{firstpage}--\pageref{lastpage}}
\maketitle


\begin{abstract}

We present photometric and spectroscopic studies of two core-collapse supernovae (SNe) 2008aq and 2019gaf in the optical wavelengths.  Light curve and spectral sequence of both the SNe are similar to those of other Type IIb SNe.  The pre-maximum spectrum of SN~2008aq showed prominent \ion{H}{$\alpha$} lines, the He lines started appearing in the near maximum spectrum. The near maximum spectrum of SN~2019gaf shows shallow \ion{H}{$\alpha$} absorption and He lines with almost similar strength. Both the SNe show transition from  hydrogen-dominated spectra to helium-dominated spectra within a month after maximum brightness. The velocity evolution of SN~2008aq matches well with those of other well-studied Type IIb SNe, while SN~2019gaf shows higher velocities. Close to maximum light, the \ion{H}{$\alpha$} and He I line velocities of SN~2019gaf are $\sim$ 2000 km sec$^{-1}$ and $\sim$ 4000 km sec$^{-1}$ higher than other well-studied Type IIb SNe. Semi-analytical modeling indicates SN~2019gaf to be a more energetic explosion with a smaller ejecta mass than SN~2008aq. The zero-age main-sequence (ZAMS) mass of the progenitor estimated using the nebular spectra of SN~2008aq ranges between 13 to 20 M$_\odot$, while for SN~2019gaf, the inferred ZAMS mass is between 13 to 25 M$_\odot$. The [\ion{O}{i}] to [\ion{Ca}{ii}] lines flux ratio favors a less massive progenitor star in a binary system for both the SNe. 

\end{abstract}




\section{Introduction}
\label{Introduction}

Core--collapse supernovae (CCSNe) represent the final stage in the evolution of massive stars ($\gtrsim 8$--$10\,M_\odot$), occurring when their core collapse under gravity  \citep{bethe1990supernova,2009ARA&A..47...63S,2012ARNPS..62..407J}. Multiple collapse channels operate for CCSNe depending on progenitor mass, structure and composition  \citep{Burbidge1957,1966ApJ...143..626C,1985ApJ...295...14B,1995ApJS..101..181W,2005NatPh...1..147W,2012ARNPS..62..407J}. Massive stars with zero-age main-sequence (ZAMS) masses above roughly $10\,M_{\odot}$ undergo successive stages of nuclear burning, synthesizing progressively heavier elements up to Fe core. Once an Fe core is produced, nuclear fusion ceases, the core then becomes unstable and collapses under its own gravity, giving rise to an iron-core collapse SN \citep{bethe1990supernova,2012ARNPS..62..407J}.  A distinct pathway is electron-capture supernovae (ECSNe) from super-AGB progenitors ($\sim$8–10$\,M_\odot$), with O-Ne-Mg cores. When the central density exceeds a critical value, electrons begin to be captured by Mg, resulting in decrease in degenerate pressure, and gravitational collapse of O-Ne-Mg core \citep{1980PASJ...32..303M,1982Natur.299..803N}. 

Type IIb SNe are a subclass of stripped-envelope (SE) CCSNe, in which the hydrogen envelope of the progenitor star is partially stripped  \citep{1988AJ.....96.1941F,2019NatAs...3..717M}. They are also designated as transitional, as their spectral evolution shows the transition from a hydrogen-rich Type II-like spectrum during the early phase to a hydrogen-poor Type Ib-like spectrum in the late post-maximum phase {\citep{1994ApJ...429..300W,1997ARA&A..35..309F,2017hsn..book..195G,2019MNRAS.485.1559P}. The light curves of Type IIb SNe are mainly powered by the radioactive decay chain of $^{56}$Ni synthesized in the explosion. A wide range of $^{56}$Ni mass (0.02 to 0.1 M$_{\odot}$) is estimated in these explosions \citep{2016MNRAS.458.2973P}. SESNe display considerable variation in their light curve shapes. Some Type IIb SNe e.g.  SNe~1993J \citep{1993ApJ...417L..71W}, 
2011dh \citep{2011ApJ...742L..18A,2013MNRAS.433....2S,2014A&A...562A..17E}, 
2011fu \citep{2015MNRAS.454...95M,10.1093/mnras/stt162}, 
2013df \citep{2016MNRAS.460.1500S}, 
2016gkg \citep{Arcavi_2017}, 
2017jgh \citep{10.1093/mnras/stab2138}, 
ZTF18aalrxas \citep{Fremling_2019}, 
2020bio \citep{Pellegrino_2023},  exhibit two peaks separated by several days in their light curves. 
The first peak is known to arise from shock cooling emission (SCE)  \citep{2012ApJ...752...78S}, and the second peak is mainly powered by the decay of $^{56}$Ni. As the shock-breakout followed by cooling dynamics, revolves around the characteristics of the outermost layers of the progenitor, early observations can yield valuable information about temperature, radius, and mass pertaining to the envelope of the progenitor or surrounding material \citep{Nakar_2014,Piro_2021}. The presence of SCE suggests that Type IIb SNe could arise from explosions of stars possessing extended outer envelopes; the duration of SCE depends on the size of this envelope \citep{2012ApJ...752...78S}. The rise times generally fall within 10 to 20 days \citep{2019MNRAS.485.1559P}, with mean peak absolute magnitudes of M$_B$ $\sim$ $-$16.99$\pm$0.45 mag for Type IIb SNe \citep{2006AJ....131.2233R,2014AJ....147..118R}. 

The spectral evolution of Type IIb SNe exhibits an interesting transition from Type II to Type Ib. Type IIb SNe display prominent hydrogen features during the early time, which follows the dominance of He lines later \citep{2000AIPC..522..123F}. 
There have been claims of hydrogen detection in some Type Ib SNe \citep{Anupama_2005,Parrent_2007,Parrent_2016}. The thin boundary between Type Ib and Type IIb SNe highly depends on the progenitor's hydrogen envelope at the time of explosion. The mass of the hydrogen envelope varies widely depending on how much mass is stripped away \citep{2010ApJ...725..940Y,2017ApJ...840...10Y,2021ApJ...913...55H,2022MNRAS.511..691G,2023ApJ...957..100G,2024A&A...685A..58E}. The post-maximum spectral evolution of Type IIb SNe, prior to the nebular phase, is dominated by helium features. During the nebular phase, oxygen, calcium, and Iron Group Elements (IGEs) are the most prominent features. 

SESNe are known to ``strip off" the outer envelopes of the progenitor before explosion \citep{2019NatAs...3..717M}. There are two scenarios proposed for a star to lose mass and end as an SESN: a massive star that loses its outer envelope due to strong stellar winds \citep{10.1093/mnras/stv2283} or a star stripping off its outer envelope because of mass transfer in a binary system \citep{1992ApJ...391..246P,10.1093/mnras/stz3224}. The most favorable progenitor scenario amongst these two is still debated, but a less massive star in a binary system is a more plausible explanation for SESNe \citep{2009ARA&A..47...63S,2011IAUS..272..474S}. Pre-explosion images are useful for gaining an insight into  the associated progenitor system. To date, there are five direct detections of the progenitor for Type IIb SNe reported and those are for SNe~1993J \citep{1994AJ....107..662A}, 2008ax \citep{2008MNRAS.391L...5C}, 2011dh \citep{2011ApJ...739L..37M,2011ApJ...741L..28V}, 2013df \citep{2014AJ....147...37V}, and 2016gkg \citep{2023MNRAS.519..471V}. For these SNe, various progenitor channels have been proposed such as a K0Ia star in a binary system with an early supergiant as companion star for SN~1993J \citep{1993ApJ...415L.103F,1994AJ....107..662A,2009Sci...324..486M}, an extended yellow supergiant for SNe~2011dh and 2013df \citep{Maund_2011,10.1093/mnras/stv2098,Folatelli_2014,2011ApJ...741L..28V,2013ApJ...772L..32V,2014AJ....147...37V,2012ApJ...757...31B}, a Wolf Rayet (WR) star in a binary system in the case of SN~2008ax \citep{2008MNRAS.391L...5C,2008MNRAS.389..955P}, and a binary system as well for SN~2016gkg \citep{2018Natur.554..497B,2023MNRAS.519..471V}. Thus, both channels, i.e., a less massive star in a binary system and a single massive star, appear equally probable as progenitors of Type IIb SNe \citep{2009ARA&A..47...63S,2011IAUS..272..474S}. 

Both SNe, 2008aq and 2019gaf, are members of the Type IIb subclass. An extensive analysis, using optical photometric and spectroscopic data, of these two SNe is presented in this paper. Section \ref{Discovery, observation, and data reduction} provides details on the discovery, the observational campaign carried out for both the SNe and the data reduction procedure. The distance adopted for both the SNe, the extinction along the line of sight to the SNe, and the epoch of explosion have been estimated in Section \ref{distance_extinction_explosion_epoch}. Section \ref{analaysis_light_curve} deals with photometric analysis. The spectroscopic analysis, evolution of the photospheric velocity, and spectral modeling are discussed in detail in Section \ref{spectroscopic_properties}. A discussion of the explosion parameters and properties of the progenitor is provided in Section \ref{explosion_parameters_progenitor_propoerties} followed by a summary in Section~\ref{summary}.

\section{Discovery, observation, and data reduction}
\label{Discovery, observation, and data reduction}

SN~2008aq (R.A.(J2000.0) 12$^h$50$^m$30.42$^s$, Dec.(J2000.0) -10$^o$52$'$01.4${''}$) was spotted by Lick Observatory Supernova Search on 27 February 2008 \citep{2008CBET.1271....1C} and classified as a Type IIb SN \citep{2014AJ....147...99M}. Optical photometric observations of SN~2008aq were initiated $\sim$3 days post-discovery using the Hanle Faint Object Spectrograph and Camera (HFOSC) mounted on the 2m Himalayan Chandra Telescope (HCT, \citealt{2010ASInC...1..193P}). Apart from the SN frames, we obtained several calibration frames, e.g., twilight flats and bias frames, during the observations. Several standard star fields from \citet{1992AJ....104..340L} were observed on four photometric nights to calibrate a sequence of secondary standards in the SN field.

Pre-processing of the photometric data, e.g. bias correction, flat fielding, etc., was done in a standard manner using various tasks available in  IRAF \footnote{IRAF stands for Image Reduction and Analysis Facility, which was distributed by the National Optical Astronomy Observatories, operated by the Association of Universities for Research in Astronomy, Inc., under a cooperative agreement with the National Science Foundation.}. Aperture photometry at an optimal aperture was used to estimate the magnitudes of the Landolt standards. The optimal aperture was determined using the aperture growth curve. The average value of atmospheric extinction for the site \citep{2008BASI...36..111S} and average color terms for the system were used to determine the photometric zero-points on individual nights. These were used to calibrate a sequence of secondary standards in the SN frame. The mean {\it BVRI} magnitudes of secondary standards in the field of SN~2008aq were estimated. As SN~2008aq occurred in the outer region of the host galaxy,  with varying background, we used profile-fitting photometry to estimate SN magnitudes. The SN magnitudes were calibrated differentially with respect to the local secondary standards in the field. The photometric observations for SN~2008aq are shown in Table \ref{tab:photometric_observational_log_2008aq}.

Spectroscopic observations of SN~2008aq were also made using HFOSC. The spectra were obtained at 13 epochs between $\sim$$-$4 days and 120 days with respect to the date of maximum brightness. The spectroscopic data reduction followed a series of standard steps, including bias correction and flat-fielding. The optimal extraction method \citep{1986PASP...98..609H} was used to extract the one-dimensional spectrum. Wavelength calibration was performed using the dispersion solution obtained from arc lamp spectra, with bright night sky emission lines employed to validate the accuracy. Small corrections were applied whenever required. The instrumental response was corrected using a response curve derived from spectro-photometric standard stars, enabling flux calibration of the spectrum. The final spectrum was generated by combining the flux-calibrated spectrum obtained  in blue and red regions, using a weighted mean. All the spectra of SN~2008aq were adjusted to the absolute flux scale based on the corresponding photometric magnitudes and corrected for redshift. Table \ref{tab:spectroscopic_observations_2008aq} lists the spectroscopic observations for SN~2008aq.

SN~2019gaf (R.A.(J2000.0) 20$^h$36$^m$55.24$^s$, Dec.(J2000.0) 02$^o$48$'$24.48${''}$) was discovered by Asteroid Terrestrial-impact Last Alert System (ATLAS, \citealt{2018PASP..130f4505T}) on May 27, 2019 \citep{2019TNSTR.862....1T}. This SN was classified as a Type IIb SN \citep{2019TNSCR.902....1S,2019TNSAN..21....1F,2019TNSCR2859....1D}. SN~2019gaf exploded in the outskirts of its anonymous host galaxy. The optical photometric campaign of SN~2019gaf began $\sim$ 8 days after discovery with the telescopes of the Las Cumbres Observatory (LCO;  \citealt{2013PASP..125.1031B}) under the Global Supernova Project (GSP) in {\it UBgVri} bands. The \texttt{lcogtsnpipe} pipeline \citep{2016MNRAS.459.3939V}  was used to estimate the SN magnitudes. Calibration of the $gri$-band instrumental magnitudes was done using the APASS catalog \citep{APASS} \footnote{\url{https://www.aavso.org/aavso-photometric-all-sky-survey-data-release-1}}. The $UBV$-band magnitudes were calibrated using the Landolt catalog \citep{1992AJ....104..340L}, constructed from standard fields observed with the same telescope and during the same nights as the SN observations. The final photometry of SN~2019gaf is presented in Table \ref{tab:photometric_observational_log_2019gaf}. Spectroscopic follow-up of SN~2019gaf started $\sim$9 days after discovery using the FLOYDS spectrograph on the 2m Faulkes Telescope North (FTN) and HCT. We have used the \texttt{floydsspec}\footnote{https://www.authorea.com/users/598/articles/6566} pipeline to perform the spectral reduction. For HCT data, we have followed a similar data reduction procedure as described for SN~2008aq. Finally, all the spectra were scaled with respect to the photometry and corrected for redshift. Details of the spectroscopic observations for SN~2019gaf are presented in Table \ref{tab:spectroscopic_observations_2019gaf}.

\begin{table*}
\caption{Optical photometric data of SN 2008aq}
\centering
\smallskip
\scriptsize
\begin{tabular}{c c c c c c c c  }
\hline \hline
Date    &   JD$^\dagger$   &   Phase$^\ddagger$ 	&   B       				&      V           	 	&  R                		& I  	                      \\
        &                  &   (Days)           	& (mag)     				& (mag)            		 & (mag)             		&(mag) 	            \\

\hline
01/03/2008  &  527.48  &  -3.92    &  16.55$\pm$0.01 & 16.09$\pm$0.01 &   15.83$\pm$0.01 &    15.86$\pm$0.02 \\
02/03/2008  &  528.41  &  -2.99    &  16.45$\pm$0.01 & 16.00$\pm$0.01 &   15.72$\pm$0.01 &    15.69$\pm$0.03 \\
04/03/2008  &  530.45  &  -0.95    &  16.35$\pm$0.01 & 15.90$\pm$0.01 &   15.62$\pm$0.01 &    15.58$\pm$0.02 \\
05/03/2008  &  531.40  &  0.0      &  16.34$\pm$0.01 & 15.87$\pm$0.01 &   15.55$\pm$0.01 &    15.54$\pm$0.01 \\
06/03/2008  &  532.34  &  0.94    &  16.34$\pm$0.01 & 15.84$\pm$0.01 &   15.55$\pm$0.01 &    15.47$\pm$0.03 \\
07/03/2008  &  533.38  &  1.98    &  16.37$\pm$0.01 & 15.84$\pm$0.01 &   15.52$\pm$0.01 &    15.47$\pm$0.02 \\
09/03/2008  &  535.43  &  4.03    &  16.51$\pm$0.01 & 15.90$\pm$0.01 &   15.53$\pm$0.01 &    15.45$\pm$0.02 \\
14/03/2008  &  540.30  &  8.90    &  17.21$\pm$0.08 & 16.20$\pm$0.01 &   15.73$\pm$0.01 &    15.57$\pm$0.02 \\
20/03/2008  &  546.29  &  14.89   &  17.96$\pm$0.04 & 16.65$\pm$0.04 &   16.10$\pm$0.02 &    15.79$\pm$0.04 \\
29/03/2008  &  555.22  &  23.82   &  18.39$\pm$0.03 & 17.07$\pm$0.01 &   16.51$\pm$0.01 &    16.07$\pm$0.03 \\
04/04/2008  &  561.22  &  29.82   &  18.47$\pm$0.02 & 17.20$\pm$0.01 &   16.69$\pm$0.01 &    16.24$\pm$0.05 \\
13/04/2008  &  570.36  &  38.96   &  18.55$\pm$0.08 & 17.36$\pm$0.02 &   16.88$\pm$0.02 &    16.46$\pm$0.05 \\   
17/04/2008  &  574.28  &  42.88   &  18.55$\pm$0.04 & 17.46$\pm$0.02 &   17.01$\pm$0.05 &    16.49$\pm$0.03 \\
28/04/2008  &  585.20  &  53.80   &  18.65$\pm$0.02 & 17.68$\pm$0.01 &   17.22$\pm$0.01 &    16.69$\pm$0.03 \\
01/05/2008  &  588.26  &  56.86   &  18.64$\pm$0.02 & 17.73$\pm$0.01 &   17.28$\pm$0.01 &    16.75$\pm$0.02 \\
06/05/2008  &  593.11  &  61.71   &  18.67$\pm$0.02 & 17.81$\pm$0.01 &   17.36$\pm$0.02 &    16.86$\pm$0.02 \\    
10/05/2008  &  597.17  &  65.77   &  18.75$\pm$0.02 & 17.89$\pm$0.02 &   17.44$\pm$0.02 &    16.91$\pm$0.02 \\     
08/06/2008  &  626.18  &  94.78   &  19.09$\pm$0.02 & 18.41$\pm$0.01 &   18.05$\pm$0.01 &    17.49$\pm$0.04 \\     
03/07/2008  &  651.16  &  119.76  &  19.29$\pm$0.02 & 18.72$\pm$0.02 &   18.32$\pm$0.04 &    17.90$\pm$0.03 \\

\hline    
\end{tabular}
\newline
$^\dagger$ JD 2,454,000+ ,
$^\ddagger$ Phase has been calculated with respect to B$_{max}$ =2454531.40  
\label{tab:photometric_observational_log_2008aq}                                                        
\end{table*}

\begin{table}
\caption{Log of spectroscopic observations for SN 2008aq from HCT}
\centering
\smallskip
\begin{tabular}{c c c }
\hline \hline
Date          & JD$^\dagger$  &Phase$^\ddagger$            \\
              &     &(Days)                            \\
\hline
01/03/2008 &     527.43&  -3.97   \\
03/03/2008 &     529.44&  -1.96     \\
04/03/2008 &     530.46&  -0.94     \\
06/03/2008 &     532.36&  0.96     \\
07/03/2008 &     533.40&  2.00     \\
12/03/2008 &     538.32&  6.80     \\
20/03/2008 &     546.30&  14.9     \\
13/04/2008 &     570.32&  38.92        \\
23/04/2008 &     580.32&  48.92        \\
28/04/2008 &     585.24&  53.84    \\
01/05/2008 &     588.28&  56.88    \\
06/05/2008 &     593.34&  61.94    \\
03/07/2008 &     651.17&  119.77    \\

\hline                          
\end{tabular}
\newline
$^\dagger$ 2454000+
$^\ddagger$ Phase is calculated with respect to  B$_{max}$= 2454531.40
\label{tab:spectroscopic_observations_2008aq}      
\end{table}

\begin{table*}
\caption{Optical photometric data of SN 2019gaf}
\centering
\smallskip
\scriptsize
\begin{tabular}{c c c c c c c c c }
\hline \hline
Date    &   JD$^\dagger$ & Phase$^\ddagger$   &  U	         &   B       				&   g           	 	 &  V                		& r   	        	 & i                   \\
        &                   &  (Days)           & (mag)     				& (mag)            		 & (mag)             		&(mag) 			&(mag)  & (mag)          \\
\hline  

2019/06/04 & 638.60 & 0.00 & 17.84$\pm$0.05 & 17.78$\pm$0.07 &  17.46$\pm$0.02        & 17.41$\pm$0.02   &  17.42$\pm$ 0.02    &  17.44$\pm$0.03    \\          
2019/06/04 & 638.60 & 0.00 & 17.75$\pm$0.14 & 17.78$\pm$0.04 &  17.46$\pm$0.02        & 17.42$\pm$0.02   &  17.46$\pm$ 0.02    &  17.44$\pm$0.03    \\    
2019/06/07 & 642.48 & 3.88 & 18.06$\pm$0.04 & 17.86$\pm$0.01 &   ----$\pm$----        & 17.34$\pm$0.01   &  -----$\pm$ ----    &  -----$\pm$----          \\    
2019/06/07 & 642.48 & 3.88 & 18.03$\pm$0.04 & 17.90$\pm$0.01 &   ----$\pm$----        & -----$\pm$----   &  -----$\pm$ ----    &  -----$\pm$----          \\    
2019/06/11 & 645.57 & 6.97 & 18.30$\pm$0.07 & 18.08$\pm$0.02 &  17.69$\pm$0.01        & 17.45$\pm$0.01   &  -----$\pm$ ----    &  -----$\pm$----          \\    
2019/06/11 & 645.57 & 6.97 & 18.15$\pm$0.05 & 18.16$\pm$0.02 &  17.67$\pm$0.01        & 17.48$\pm$0.01   &  -----$\pm$ ----    &  -----$\pm$----          \\    
2019/06/13 & 647.54 & 8.94 & 18.51$\pm$0.06 & 18.16$\pm$0.02 &  17.78$\pm$0.01        & 17.45$\pm$0.02   &  17.26$\pm$ 0.01    &  -----$\pm$----          \\    
2019/06/13 & 647.54 & 8.94 & 18.35$\pm$0.06 & 18.21$\pm$0.02 &  17.81$\pm$0.01        & 17.43$\pm$0.01   &  17.26$\pm$ 0.01    &  -----$\pm$----          \\    
2019/06/15 & 650.47 & 11.87 & 18.61$\pm$0.10 & 18.38$\pm$0.05 &  17.98$\pm$0.02        & 17.53$\pm$0.03   &  17.33$\pm$ 0.02    &  17.19$\pm$0.02    \\    
2019/06/15 & 650.48 & 11.88 & -----$\pm$---  & 18.41$\pm$0.05 &  17.96$\pm$0.03        & 17.55$\pm$0.03   &  -----$\pm$-----    &  17.21$\pm$0.02   \\    
2019/06/20 & 654.62 & 16.02 & -----$\pm$---  & 18.88$\pm$0.08 &  18.32$\pm$0.05        & 17.76$\pm$0.03   &  17.54$\pm$ 0.03    &  17.50$\pm$0.07    \\    
2019/06/20 & 654.62 & 16.02 & -----$\pm$---  & 19.04$\pm$0.07 &  18.38$\pm$0.04        & 17.74$\pm$0.04   &  17.58$\pm$ 0.04    &  -----$\pm$----        \\    
2019/06/23 & 657.89 & 19.29 & -----$\pm$---  & 19.14$\pm$0.04 &  18.70$\pm$0.03        & 18.02$\pm$0.02   &  17.69$\pm$ 0.02    &  17.53$\pm$0.02    \\    
2019/06/23 & 657.89 & 19.29 & -----$\pm$---  & 19.29$\pm$0.04 &  18.64$\pm$0.03        & 18.07$\pm$0.02   &  17.73$\pm$ 0.01    &  17.52$\pm$0.02    \\    
2019/06/27 & 662.22 & 23.62 & -----$\pm$---  & 19.70$\pm$0.05 &  19.03$\pm$0.02        & 18.40$\pm$0.02   &  17.95$\pm$ 0.01    &  17.67$\pm$0.02    \\    
2019/06/27 & 662.22 & 23.62 & -----$\pm$---  & 19.63$\pm$0.05 &  19.05$\pm$0.02        & 18.35$\pm$0.02   &  17.96$\pm$ 0.01    &  17.66$\pm$0.02    \\    
2019/07/01 & 666.16 & 27.56 & -----$\pm$---  & 19.94$\pm$0.05 &  19.23$\pm$0.05        & 18.56$\pm$0.03   &  18.11$\pm$ 0.07    &  17.97$\pm$0.06    \\    
2019/07/01 & 666.16 & 27.56 & -----$\pm$---  & 20.01$\pm$0.06 &  19.27$\pm$0.06        & 18.51$\pm$0.05   &  18.20$\pm$ 0.03    &  18.15$\pm$0.08    \\    
2019/07/09 & 673.52 & 34.92 & -----$\pm$---  & 20.05$\pm$0.07 &   ----$\pm$----        &  ----$\pm$----   &   ----$\pm$ ----    &   ----$\pm$----          \\ 
2019/07/10 & 674.83 & 36.23 & -----$\pm$---  & -----$\pm$---- &  19.22$\pm$0.06        & -----$\pm$----   &   ----$\pm$ ----    &  18.24$\pm$0.10     \\ 
2019/07/10 & 674.84 & 36.24 & -----$\pm$---  & -----$\pm$---- &  -----$\pm$----        & -----$\pm$----   &   ----$\pm$-----    &  18.30$\pm$0.07     \\  
2019/07/13 & 678.40 & 39.80 & -----$\pm$---  & -----$\pm$---- &  19.45$\pm$0.09        & 19.11$\pm$0.10   &   ----$\pm$ ----    &   ----$\pm$----          \\
2019/07/13 & 678.40 & 39.80 & -----$\pm$---  & -----$\pm$---- &  19.47$\pm$0.09        & 19.09$\pm$0.10   &  18.65$\pm$ 0.05    &   ----$\pm$----          \\
2019/07/13 & 678.42 & 39.82 & -----$\pm$---  & -----$\pm$---- &  19.53$\pm$0.07        & 19.29$\pm$0.08   &  18.65$\pm$ 0.05    &  18.42$\pm$0.06     \\
2019/07/13 & 678.42 & 39.82 & -----$\pm$---  & -----$\pm$---- &  -----$\pm$----        & 19.14$\pm$0.09   &  18.68$\pm$ 0.05    &  18.47$\pm$0.05     \\ 
2019/07/22 & 687.08 & 48.48 & -----$\pm$---  & 20.31$\pm$0.10 &  19.84$\pm$0.05        & 19.34$\pm$0.04   &  18.91$\pm$ 0.03    &  18.56$\pm$0.04     \\    
2019/07/22 & 687.08 & 48.48 & -----$\pm$---  & 20.44$\pm$0.11 &  19.70$\pm$0.05        & 19.29$\pm$0.04   &  18.89$\pm$ 0.03    &  18.52$\pm$0.03     \\    
2019/07/31 & 695.72 & 57.12 & -----$\pm$---  & 20.27$\pm$0.04 &  19.71$\pm$0.02        & 19.18$\pm$0.03   &  -----$\pm$ ----    &  -----$\pm$----          \\    
2019/07/31 & 695.72 & 57.12 & -----$\pm$---  & 20.23$\pm$0.03 &  19.72$\pm$0.02        & 19.20$\pm$0.02   &  -----$\pm$ ----    &  -----$\pm$----          \\    
2019/08/09 & 705.45 & 66.85 & -----$\pm$---  & 20.35$\pm$0.07 &  19.74$\pm$0.03        & -----$\pm$----   &  -----$\pm$ ----    &  -----$\pm$----          \\    
2019/08/17 & 713.47 & 74.87 & -----$\pm$---  & -----$\pm$---- &  -----$\pm$----        & -----$\pm$----   &  -----$\pm$-----    &  18.79$\pm$0.06     \\
2019/08/17 & 713.48 & 74.88 & -----$\pm$---  & -----$\pm$---- &  -----$\pm$----        & -----$\pm$----   &  -----$\pm$-----    &  18.94$\pm$0.08     \\
2019/08/25 & 721.38 & 82.78 & -----$\pm$---  & 20.41$\pm$0.05 &  20.01$\pm$0.03        & 19.58$\pm$0.03   &  19.19$\pm$ 0.02    &  -----$\pm$----          \\  
2019/08/25 & 721.39 & 82.79 & -----$\pm$---  & -----$\pm$---- &  -----$\pm$----        & 19.58$\pm$0.03   &  19.12$\pm$ 0.02    &  -----$\pm$----          \\  
2019/09/26 & 752.70 & 114.10 & -----$\pm$---  & 20.88$\pm$0.06 &  -----$\pm$----        & 20.18$\pm$0.04   &  19.65$\pm$ 0.04    &  19.77$\pm$0.07      \\    
2019/09/26 & 752.70 & 114.10 & -----$\pm$---  & 21.02$\pm$0.07 &  -----$\pm$----        & -----$\pm$----   &  -----$\pm$ ----    &  -----$\pm$----          \\    
2019/10/22 & 778.88 & 140.20 & -----$\pm$---  & 20.68$\pm$0.07 &  20.40$\pm$0.04        & 20.23$\pm$0.05   &  -----$\pm$ ----    &  -----$\pm$----          \\   
2019/10/30 & 786.64 & 148.00 & -----$\pm$---  & -----$\pm$---- &  20.68$\pm$0.04        & 20.72$\pm$0.05   &  -----$\pm$ ----    &  20.16$\pm$0.06    \\
2019/10/30 & 786.65 & 148.00 & -----$\pm$---  & -----$\pm$---- &  -----$\pm$----        & -----$\pm$----   &  20.15$\pm$ 0.04    &  20.24$\pm$0.06    \\

\hline    
\end{tabular}
\newline
$^\dagger$ JD 2,458,000+ ,
$^\ddagger$ Phase has been calculated with respect to B$_{max}$ = 2458638.60  
\label{tab:photometric_observational_log_2019gaf}                                                        
\end{table*}

\begin{table}
\caption{Log of spectroscopic observations for SN 2019gaf}
\centering
\smallskip
\begin{tabular}{c c c c }
\hline \hline
Date          & JD$^\dagger$  &Phase$^\ddagger$   & Telescope         \\
              &     &(Days)                            \\
\hline
05/06/2019 &     640.43 &  1.83 &  LCO \\
06/06/2019 &     641.27 &  2.67 &  HCT  \\
09/06/2019 &     644.06 &  5.46 &  LCO  \\
13/06/2019 &     648.01 &  9.41 &  HCT  \\
13/06/2019 &     648.38 &  9.78 &  LCO  \\
25/06/2019 &     660.00 &  21.41 &  LCO   \\
02/07/2019 &     667.35 &  28.75  & HCT      \\
10/07/2019 &     674.98 &  36.38 & LCO      \\
05/09/2019 &     732.21 &  93.61 & HCT  \\
23/09/2019 &     750.21 &  111.62 & HCT   \\

\hline                          
\end{tabular}
\newline
$^\dagger$ 2458000+
$^\ddagger$ Phase is calculated with respect to  B$_{max}$= 2458638.60	
\label{tab:spectroscopic_observations_2019gaf}      
\end{table}

\section{Distance, extinction, and explosion epoch}
\label{distance_extinction_explosion_epoch}
 
SN~2008aq exploded in galaxy MCG -2-33-20 at a redshift of 0.008 \citep{2019MNRAS.482.1545S}. There are several distance estimates ranging from 30.8 to 33.8 Mpc for MCG -2-33-20 \citep{2013AJ....146...86T,2014MNRAS.444..527S,2016AJ....152...50T}. We have used the most recent distance measurement (32.8 Mpc, \citealt{2016AJ....152...50T}) and scaled it to $H_0$ = 73 km s$^{-1}$ Mpc$^{-1}$. The measured distance for SN~2008aq is 33.70 Mpc. SN~2008aq is located at the outskirt of the host galaxy MCG -2-33-20, hence negligible host extinction is expected, which is further supported by the absence of NaID line in the spectral evolution. The reddening within the Milky Way towards MCG -2-33-20 is  $E(B-V)$ = 0.04 mag \citep{2011ApJ...737..103S}, which corresponds to A$_{V}$ = 0.122 mag  assuming R$_{V}$ = 3.1. Our calculation is consistent with the extinction reported for SN 2008aq in \cite{2014ApJS..213...19B}.
\cite{2016MNRAS.461.2019S} presented spectropolarimetry of SN~2008aq and estimated that SN~2008aq was discovered $\sim$ 8 days before maximum in {\it V} band using the light curves of SN 2008aq published in \cite{2014ApJS..213...19B}. Assuming a {\it V} band rise time of $\sim$ 20 days \citep{2006AJ....131.2233R}, they estimated explosion date as February 16, 2008 (JD = 2454512.75). The {\it V} band maximum estimated with our data falls on March 06, 2008 (JD = 2454532.33$\pm$1), which corresponds to February 15, 2008 as the explosion date. Our estimates are consistent with those published in \cite{2016MNRAS.461.2019S}. Throughout this paper, we have used February 16, 2008, as the explosion date for SN~2008aq.

The redshift of SN~2019gaf is estimated to be 0.02, based on the narrow \ion{H}{$\alpha$} emission line prominently visible in the $\sim$36 day SN spectrum. Assuming $H_0$ = 73 km s$^{-1}$ Mpc$^{-1}$, $\Omega_m$ = 0.27, and $\Omega_v$ = 0.73, the corresponding luminosity distance is calculated to be 83.4 Mpc. There is no prominent NaID line in the spectral sequence of SN~2019gaf. Hence, we use only Galactic extinction as the total extinction to the line of sight, which is {\it E(B-V)} = 0.092 mag \citep{2011ApJ...737..103S} and A$_{V}$ = 0.284 using R$_{V}$ = 3.1.  
Considering the last non-detection (JD 2458629.061 $>18.86~$mag) and first detection (JD 2458631.04, 18.64$\pm$0.07~mag) of SN~2019gaf in ATLAS-$o$ band, the explosion epoch of SN~2019gaf was constrained to JD$\sim$2458630$\pm$1, which we consider throughout for further analysis.

\section{Analysis of the light curve}
\label{analaysis_light_curve}

\subsection{Light curve and color curve}
\label{light_curve_color_curve}

Figures~\ref{fig:light_curve_2008aq} and \ref{fig:light_curve_2019gaf} present the light curve evolution of SNe~2008aq and 2019gaf, respectively. The light curve of SN~2008aq is well sampled around the peak, whereas for SN~2019gaf, the peak in the bluer bands was missed in our observing campaign. The shock cooling signature is not seen in the light curve of both the objects. We used a polynomial fit to estimate the peak magnitude and epoch of maximum light; the estimated parameters are presented in Table \ref{tab:decay_rate_08aq_219gaf}. \cite{2014ApJS..213...19B} reported the {\it UBVri} photometry of SN~2008aq, which are in agreement with the derived magnitudes presented here. 

\begin{figure}
	\begin{center}
		\includegraphics[width=\columnwidth]{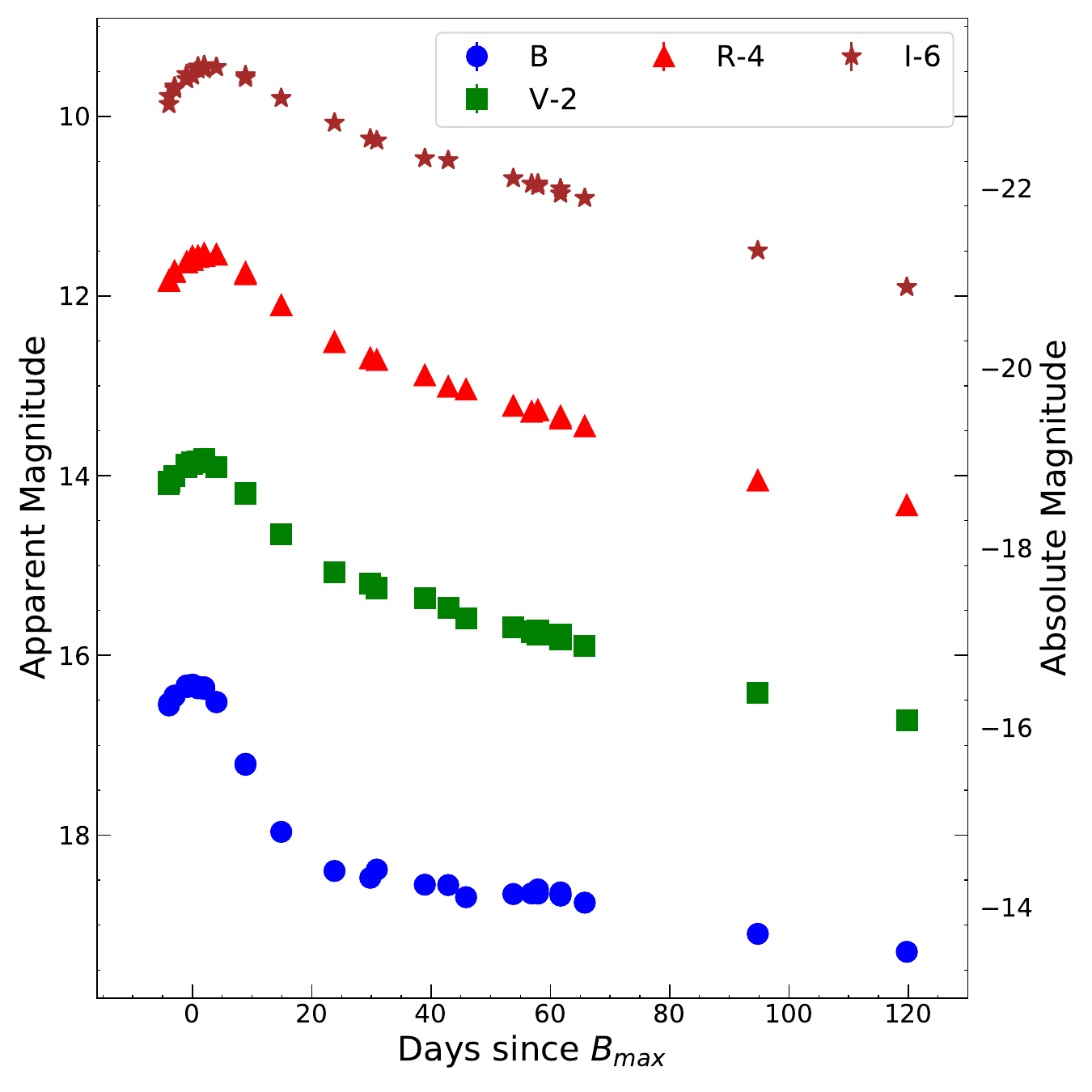}
	\end{center}
	\caption{ Light curve evolution of SN~2008aq in {\it BVRI} bands.}
	\label{fig:light_curve_2008aq}
\end{figure}

\begin{figure}
	\begin{center}
		\includegraphics[width=\columnwidth]{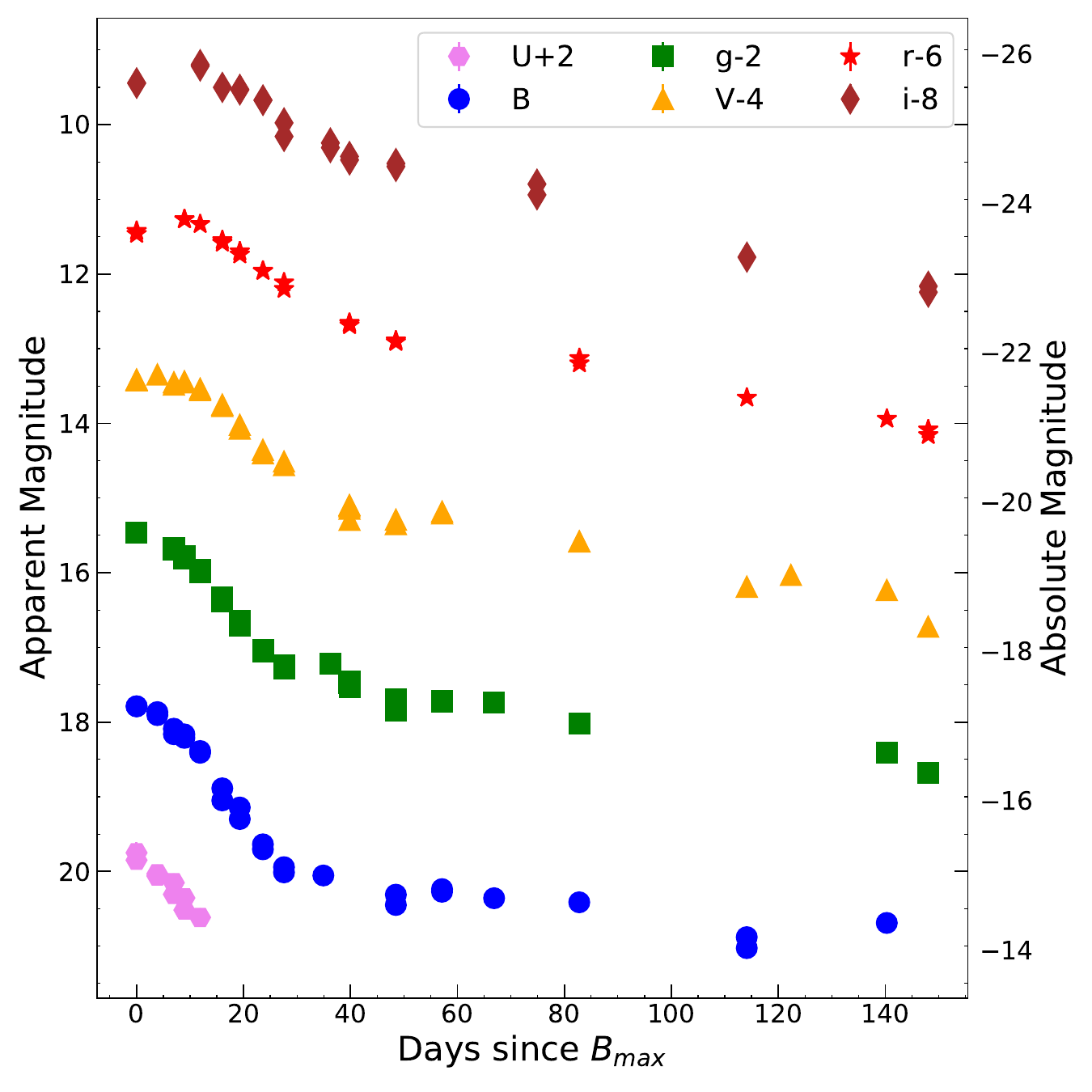}
	\end{center}
	\caption{Light curve evolution of SN~2019gaf as observed in the {\it UBgVri} bands.}
	\label{fig:light_curve_2019gaf}
\end{figure}

\begin{table*}
\caption{Light curve parameters of SNe 2008aq and 2019gaf}
\centering
\smallskip
\begin{tabular}{l  c c c c c}
\hline
SN 2008aq                          & B band           & g band           & V band            & r/R band              & i/I band  \\
\hline
JD of maximum light (2454000+)      & 531.40$\pm$0.5    &  --  &  532.33$\pm$0.5  & 533.38$\pm$0.5      &535.43$\pm$0.5     \\
Magnitude at maximum (mag)          & 16.34$\pm$0.01    &  --  &  15.84$\pm$0.01  & 15.53$\pm$0.02      & 15.45$\pm$0.02    \\
Absolute magnitude at maximum (mag) & $-$16.46$\pm$0.01 & --   & $-$16.95$\pm$0.01 & $-$17.25$\pm$0.01   & $-$17.30$\pm$0.03      \\

\hline
$\Delta$m$_{15}$( mag)                  & 1.62$\pm$0.04  & --       & 0.86$\pm$0.05     & 0.69$\pm$0.04    & 0.44$\pm$0.07   \\ 
\hline\hline
SN 2019gaf                          &          &            &          &              &  \\
\hline
JD of maximum light (2458000+)      & 638.60$\pm$2.0  & 638.60$\pm$2  & 642.49$\pm$1    & 647.55$\pm$0.5      &650.5$\pm$0.5     \\
Magnitude at maximum (mag)          & 17.79$\pm$0.08  & 17.46$\pm$0.02  & 17.35$\pm$0.02  & 17.27$\pm$0.02     & 17.19$\pm$0.03    \\
Absolute magnitude at maximum (mag) & $-$17.26$\pm$0.08 & $-$17.58$\pm$0.21  & $-$17.77$\pm$0.02 & $-$17.78$\pm$0.02     & $-$17.85$\pm$0.03    \\

\hline
$\Delta$m$_{15}$( mag) & 0.92$\pm$0.1    & 0.64$\pm$0.08     & 0.70$\pm$0.03      & 0.70$\pm$0.02    & 0.70$\pm$0.08   \\                       
\hline

\end{tabular}
\newline
$^\ddagger$with respect to B$_{max}$= 2454531.40 for SN 2008aq and B$_{max}$= 2458638.60 for SN 2019gaf
\label{tab:decay_rate_08aq_219gaf}      
\end{table*}

Figure~\ref{fig:comp_light_curve} shows the light curves evolution of SNe~2008aq and 2019gaf in {\it BVRI} bands and their comparison with a few other Type IIb - SNe~2008ax \citep{2011MNRAS.413.2140T}, 2010as \citep{2014ApJ...792....7F}, 2011dh \citep{2014A&A...562A..17E,2015A&A...580A.142E}, 2013df \citep{2016MNRAS.460.1500S,2019MNRAS.482.1545S}, and 2015as \citep{2018MNRAS.476.3611G}. The {\it r} and {\it i} band magnitudes of SN~2019gaf were converted to {\it R} and {\it I} band magnitudes using the expressions provided in \cite{2006A&A...460..339J}. The magnitudes for each SN are adjusted to their peak values in the respective bands. The figure shows that in {\it B} and {\it V} bands, up to 20 days since maximum,   SNe~2008aq and 2019gaf share similar light curve evolution with each other and with SN~2015as. In the {\it R} and {\it I} bands, the light curve evolution of SNe~2008aq and 2019gaf is similar to other SNe compared in Figure \ref{fig:comp_light_curve}. 

\begin{figure}
	\begin{center}
		\includegraphics[width=0.9\columnwidth]{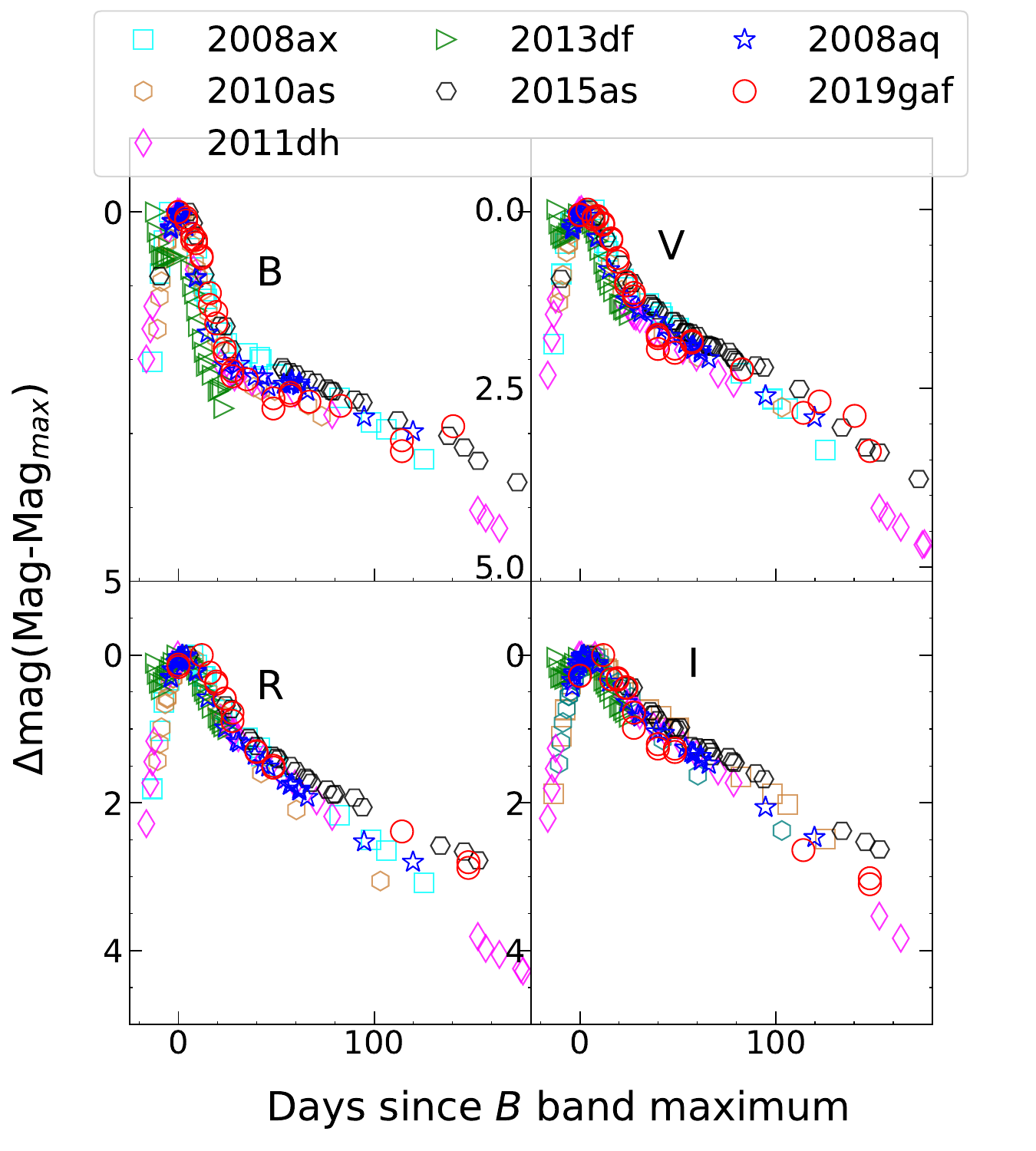}
	\end{center}
	\caption{ {\it BVRI} light curves for SNe~2008aq and 2019gaf, compared with those of other Type IIb SNe.}
	\label{fig:comp_light_curve}
\end{figure}

The evolution of {\it B-V}, {\it V-I}, {\it V-R}, and {\it R-I} colors of SNe~2008aq and 2019gaf are shown in Figure \ref{fig:color_curve}. We have also compared it with other well-studied Type IIb SNe in the same figure. Color evolution of SNe 2008aq and 2019gaf appear consistent with the patterns exhibited by other Type IIb SNe used for comparison.

\begin{figure}
	\begin{center}
		\includegraphics[width=\columnwidth]{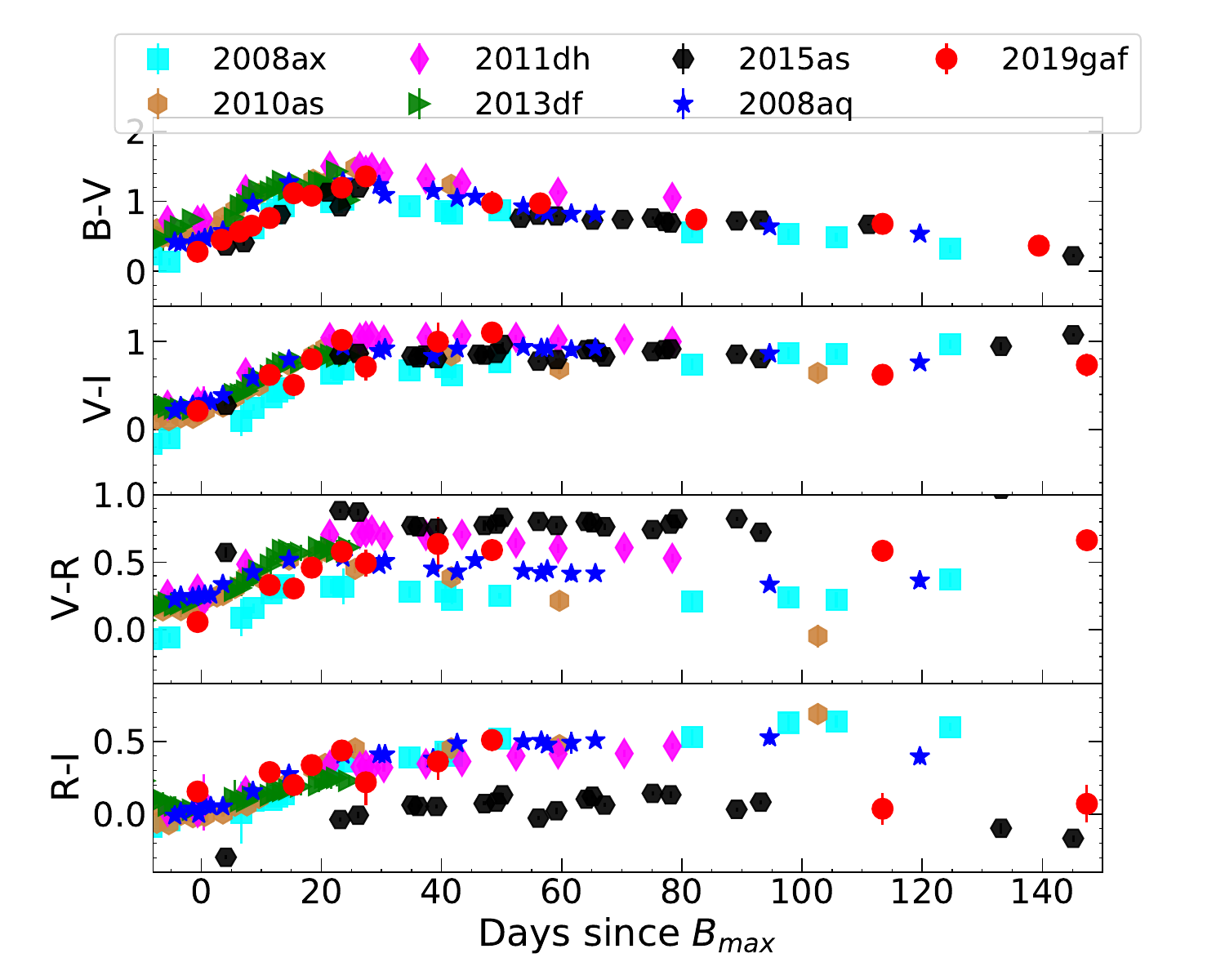}
	\end{center}
	\caption{Color evolution of SNe 2008aq and 2019gaf compared with color evolution of other well-studied Type IIb SNe.}
	\label{fig:color_curve}
\end{figure}

\subsection{Bolometric light curve}
\label{bolometric_light_curve}

Figure \ref{fig:bolometric_plot} displays the evolution of pseudo-bolometric light curves of SNe~2008aq and 2019gaf and their comparison with a few Type IIb SNe. We have used Superbol code \citep{2018RNAAS...2..230N} to calculate the pseudo-bolometric luminosities of SNe~2008aq and 2019gaf. The pseudo-bolometric light curves of other SNe in Figure \ref{fig:bolometric_plot} are constructed in a similar manner. The figure shows that the pseudo-bolometric luminosity evolution of SN~2008aq lies between SNe~2011dh and 2013df. The pseudo-bolometric light curve evolution of SN~2019gaf lies on the brighter side and matches well with SNe~2008ax and 2010as. Peak pseudo-bolometric luminosities of SNe~2008aq and 2019gaf are 41.92$\pm$0.02 and 42.17$\pm$0.01 erg s$^{-1}$, respectively. 

\begin{figure}
	\begin{center}
		\includegraphics[width=\columnwidth]{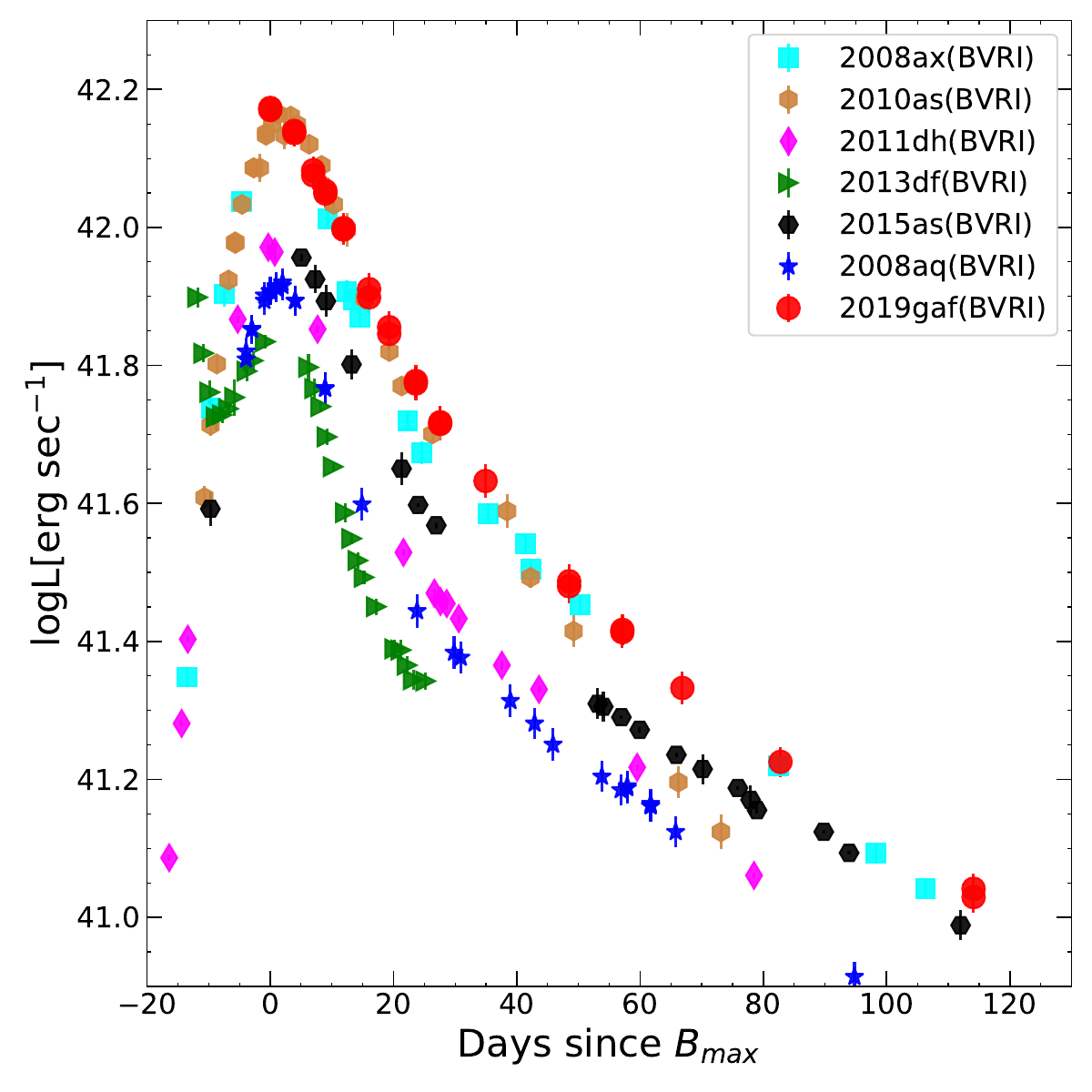}
	\end{center}
	\caption{Evolution of the pseudo-bolometric luminosity for SNe~2008aq and 2019gaf, along with a comparison to other Type IIb SNe.}
	\label{fig:bolometric_plot}
\end{figure}

\section{Spectroscopic properties}
\label{spectroscopic_properties}

\subsection{Spectral features and comparison with other Type IIb SNe }
\label{spectral_features_comparison_other_SNe}

Figures \ref{fig:spectra_2008aq} and \ref{fig:spectra_2019gaf} display the spectral time series of SNe~2008aq and 2019gaf, respectively. Spectral evolution of SN~2008aq covers phase $\sim$ 4 days before to $\sim$ 120 days after the maximum light. P-Cygni profile of \ion{H}{$\alpha$} and \ion{H}{$\beta$} are clearly seen in the early spectral sequence. The \ion{Fe}{ii} lines near 5000 \AA~ are also present in the spectrum obtained at $\sim -4$ days. The \ion{He}{i} line at 5876 \AA~ is seen in the early spectrum. The flat-topped profile of the \ion{H}{$\alpha$} emission observed in the spectral evolution of SN~2008aq is a result of the weak blue-shifted absorption feature of \ion{He}{i} line at 6678 \AA \ \citep{2016MNRAS.461.2019S}. Similar to other Type IIb SNe, SN~2008aq undergoes a transition phase approximately 15 days after maximum brightness, characterized by the dominance of He lines. Around the same time, emission due to \ion{Ca}{ii} NIR triplet also starts appearing in the spectrum of SN~2008aq. The last four spectra at $\sim$ 54, 57, 62, and 120 days exhibit signatures of [\ion{Ca}{ii}] emission lines. In the last observed spectrum at $\sim$120 days, the [\ion{O}{i}] lines at 6300 and 6364 \AA~ are also clearly present, indicating that the SN has entered the nebular phase.   

\begin{figure}
	\begin{center}
		\includegraphics[width=\columnwidth]{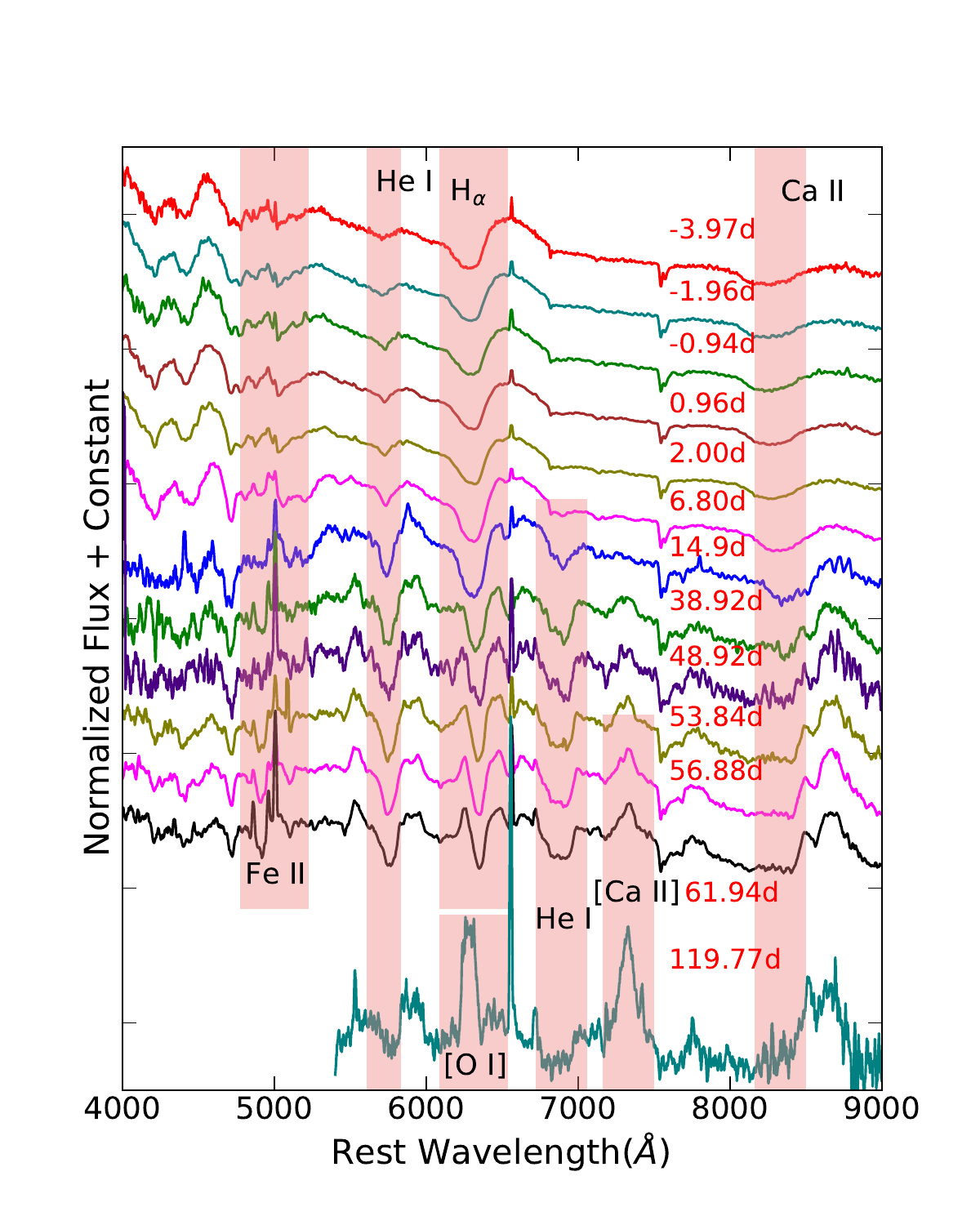}
	\end{center}
	\caption{Spectral evolution of SN~2008aq spanning from $-$3.9 days to $+$119.7 days with respect to  maximum light.}
	\label{fig:spectra_2008aq}
\end{figure}

Spectral coverage of SN~2019gaf from maximum to $\sim$+111.6 days  is presented in Figure \ref{fig:spectra_2019gaf}. The first spectrum of SN~2019gaf, prior to its peak brightness, is obtained from the Transient Name Server \citep{2019TNSAN..21....1F}.  
Similar to SN~2008aq, SN~2019gaf also shows typical signatures of a Type IIb SN. SN~2019gaf exhibits weak hydrogen features, implying that the progenitor star preserved a thin hydrogen envelope prior to the explosion. The transition from hydrogen dominated  to helium dominated spectrum in SN~2019gaf occurs  approximately 21 days after maximum brightness. The two late phase spectra of SN~2019gaf,  at 93 and 111 days post-maximum, show the emergence of  [\ion{O}{i}] and [\ion{Ca}{ii}] emission lines, which mark the transition to the nebular phase. 

\begin{figure}
	\begin{center}
		\includegraphics[width=\columnwidth]{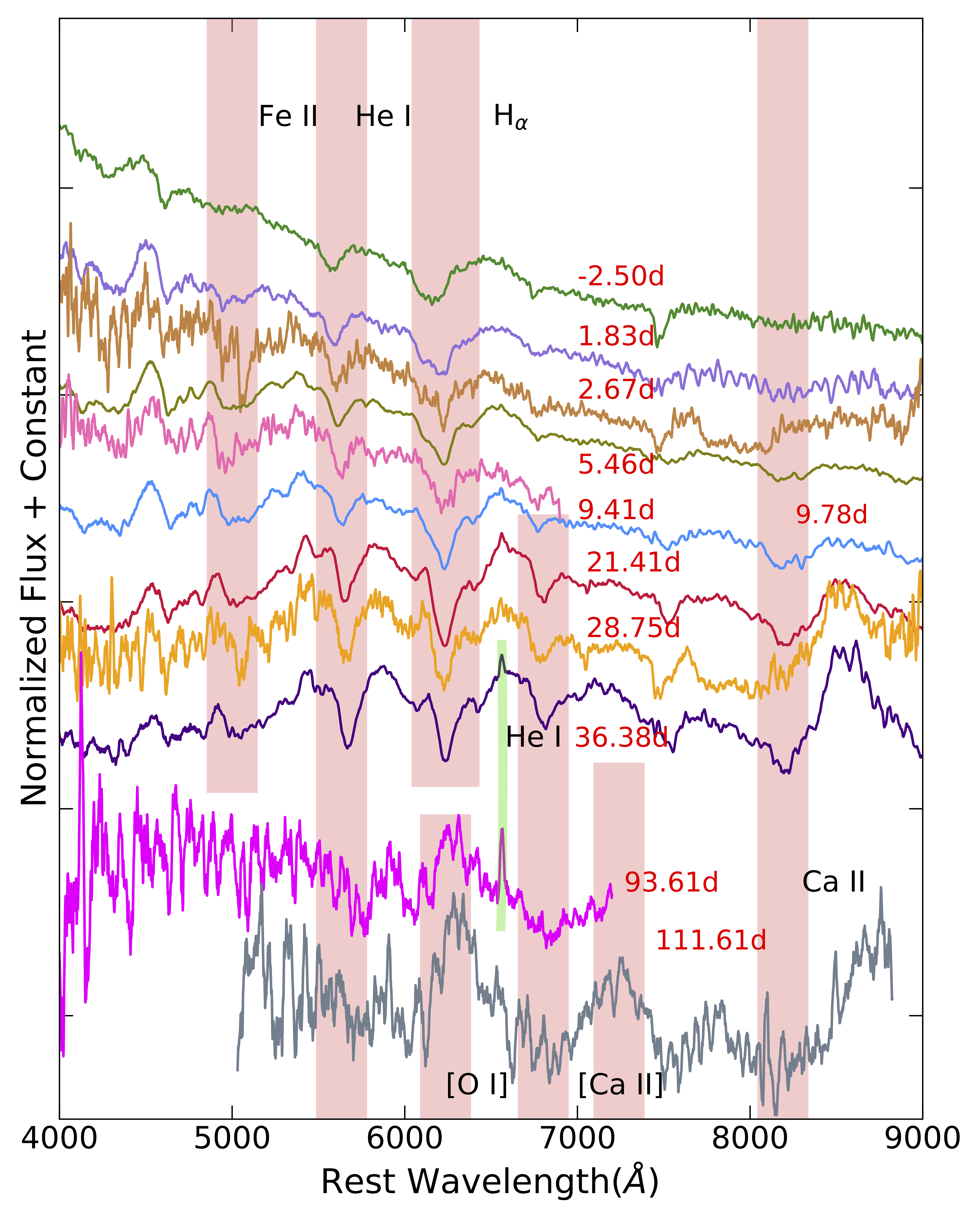}
	\end{center}
	\caption{This figure shows the spectral time series for SN~2019gaf, extending from $-$2.5 days to 111.6 days after maximum light. The narrow \ion{H}{$\alpha$} region used to estimate the redshift is indicated by the shaded green bar in the plot.}
	\label{fig:spectra_2019gaf}
\end{figure}

We have compared the spectra of SNe~2008aq and 2019gaf with a few other well-studied Type IIb SNe, namely, SNe~2008ax \citep{2011MNRAS.413.2140T,2014AJ....147...99M}, 2010as \citep{2014ApJ...792....7F}, 2011dh \citep{2013MNRAS.433....2S,2014A&A...562A..17E,2015A&A...580A.142E}, 2013df \citep{2016MNRAS.460.1500S,2019MNRAS.482.1545S}, and 2015as \citep{2018MNRAS.476.3611G}. Figures \ref{fig:near_max_comp}, \ref{fig:post_max_comp}, and \ref{fig:nebular_phase_comp} present the comparison of spectral features of SNe~2008aq and 2019gaf with other Type IIb SNe in the near-maximum, post-maximum, and nebular phases. Comparison of the near maximum spectra of SNe~2008aq and 2019gaf show that \ion{H}{$\alpha$}, \ion{H}{$\beta$}, \ion{He}{i} line at 5876 \AA, and \ion{Fe}{ii} lines near 5000 \AA~ are clearly visible in all the SNe (see Figure \ref{fig:near_max_comp}). 

The \ion{He}{i} 5876 \AA \  absorption in SN~2008aq appears weaker than in other SNe. The \ion{Ca}{ii} NIR feature is weak in SN~2019gaf as compared to other SNe. Post-maximum spectral evolution ($\sim$one month after maximum) of SNe~2008aq and 2019gaf is compared with other SNe in Figure \ref{fig:post_max_comp}. During this phase, these SNe showed transition from hydrogen-dominated to helium-dominated spectra, however, hydrogen lines are still seen in spectra of all the objects. The \ion{Fe}{ii} lines near 5000 \AA, \ion{He}{i} lines at 5876 \AA, 6678 \AA, and a clear hint of He line at 7065 \AA, are seen in all the spectra. The emission component of the \ion{Ca}{ii} NIR triplet is well developed. The comparison of nebular phase spectra of SNe~2008aq and 2019gaf is shown in Figure \ref{fig:nebular_phase_comp}. At this stage, the spectra of Type IIb SNe are mostly dominated by the \ion{Mg}{i}] line at 4571 \AA, \ion{Fe}{ii} lines near 5000 \AA, [\ion{O}{i}] lines at 6300, 6364 \AA, [\ion{Ca}{ii}] lines at 7291 and 7324 \AA, and \ion{Ca}{ii} NIR triplet, with varying strengths. Although the wavelength coverage of the nebular phase spectra of SNe 2008aq and 2019gaf is limited and the spectra have a poor signal-to-noise ratio, the \ion{Na}{i}, [\ion{O}{i}], [\ion{Ca}{ii}] and \ion{Ca}{ii} NIR triplet emission lines are clearly detected. It is worth noting that the [\ion{O}{i}] and [\ion{Ca}{ii}] lines in the last available spectra of SN~2019gaf are significantly broader than the other objects used for comparison. 

\begin{figure}
	\begin{center}
		\includegraphics[width=\columnwidth]{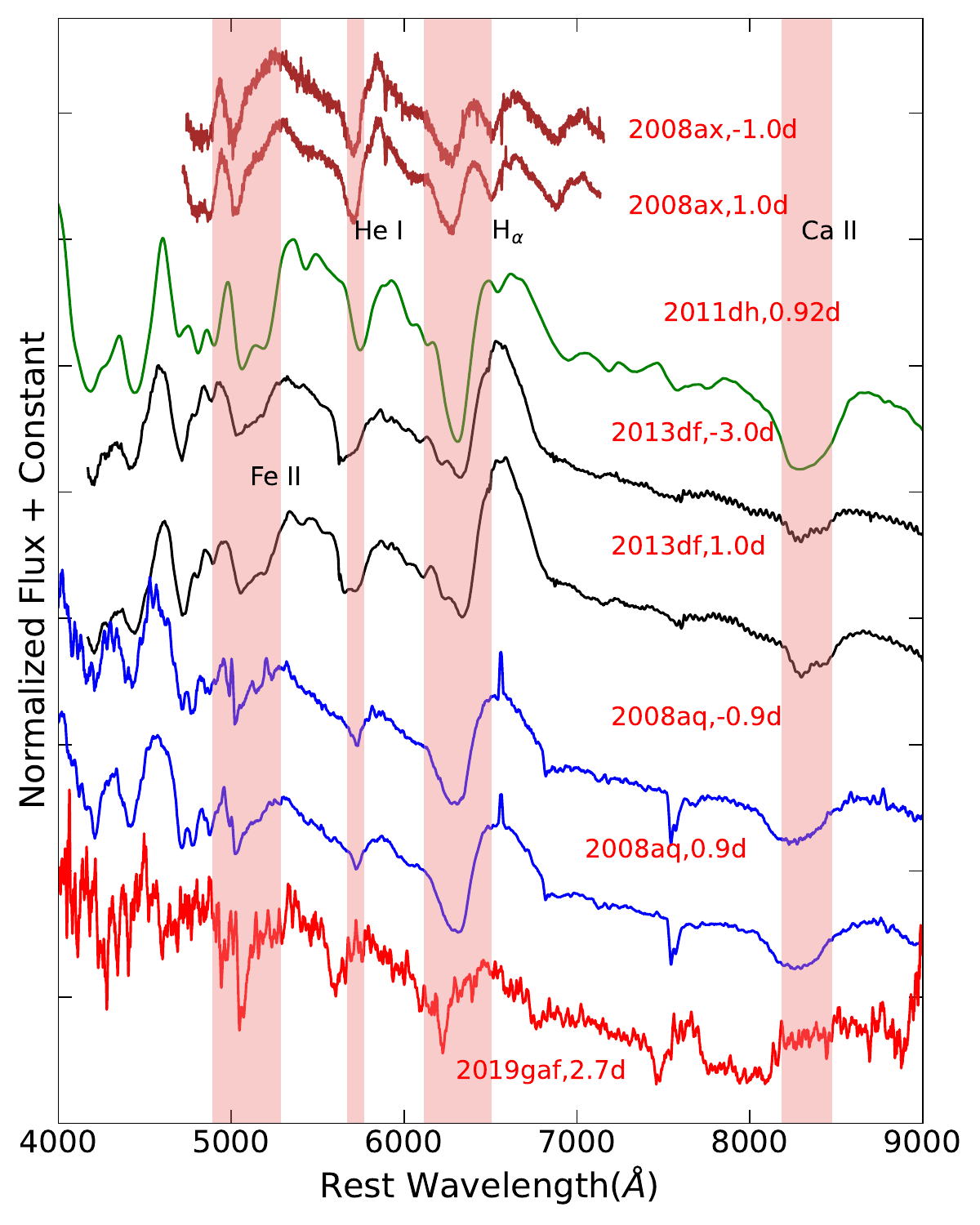}
	\end{center}
	\caption{Comparison of the near maximum spectral features of SNe~2008aq and 2019gaf with those of other well-studied Type IIb SNe at similar epochs.}
	\label{fig:near_max_comp}
\end{figure}

\begin{figure}
	\begin{center}
		\includegraphics[width=\columnwidth]{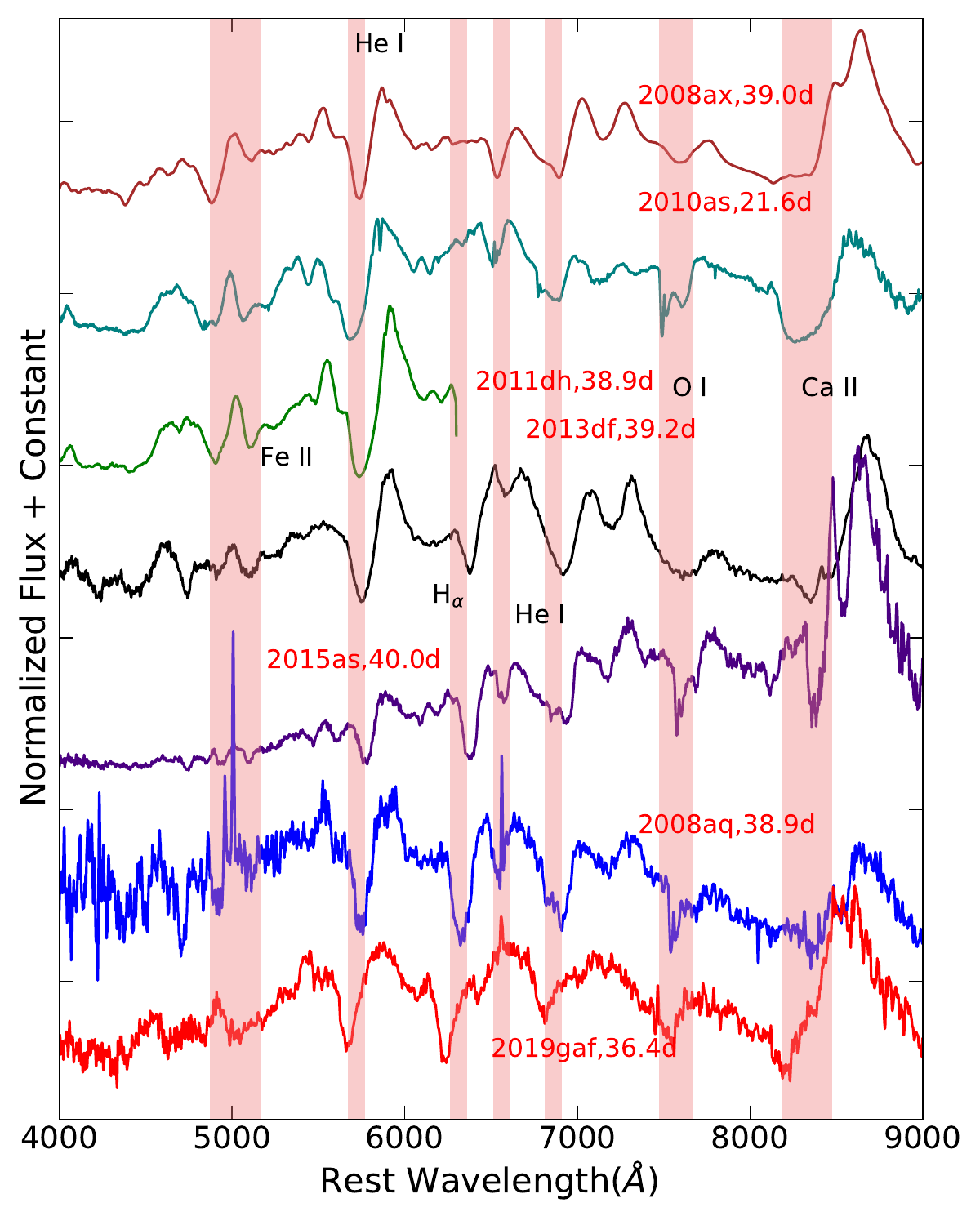}
	\end{center}
	\caption{This figure shows a comparison of the spectral features observed post-maximum light.}
	\label{fig:post_max_comp}
\end{figure}

\begin{figure}
	\begin{center}
		\includegraphics[width=\columnwidth]{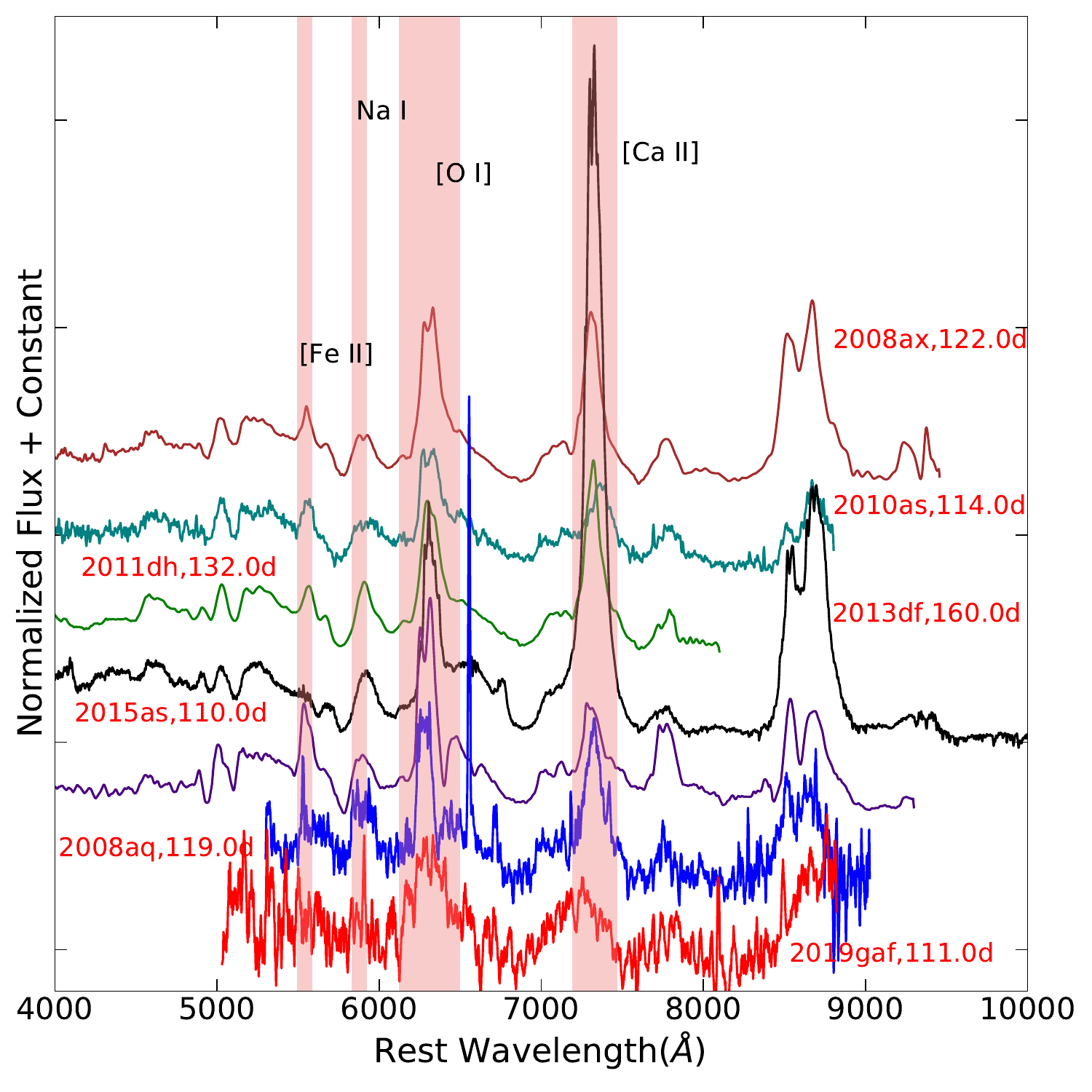}
	\end{center}
	\caption{This figure compares the nebular phase spectra of SNe 2008aq and 2019gaf with those of other Type IIb SNe.}
	\label{fig:nebular_phase_comp}
\end{figure}

\subsection{Spectral modeling}
\label{spectral_modelling}

The spectra of SN~2008aq at ten epochs, from $-$3.9 to 53.8 days relative to the $B$-band peak brightness, are modeled using the spectrum synthesis code \texttt{SYNAPPS} \citep{2011PASP..123..237T}, and are shown in Figure \ref{fig:synapps_2008aq}. The photospheric velocity in the best-fit model drops from 12,200 km s$^{-1}$ at $-$3.9 days to 7900 km s$^{-1}$ at 53.8 days. The outer ejecta velocity used in the modeling is 30,000 km s$^{-1}$. The photospheric temperature evolves from 11,200 K at $-$3.9 days to 6900 K at 53.8 days. The chemical species used in the modeling are H, He, O, Na, Ca, Mg, and Fe; specifically, \ion{H}{i}, \ion{He}{i}, \ion{O}{i}, \ion{Na}{i}, \ion{Ca}{ii}, \ion{Mg}{ii}, and \ion{Fe}{ii} ionization states were used. Most of the features are well-reproduced in the early phases, while the fit starts to deteriorate from 38.9 days when the SN enters the nebular phase. This is expected as the LTE assumption of \texttt{SYNAPPS} does not remain valid in the nebular phase. 

Similarly, we modeled four spectra of SN~2019gaf, from 1.8 to 21.4 days relative to the $B$-band peak brightness, using \texttt{SYNAPPS} and the same chemical species and outer ejecta velocity as SN~2008aq. The spectra and corresponding models are shown in Figure \ref{fig:synapps_2019gaf}. In the best-fit model, the photospheric velocity decreases from 15,000 km s$^{-1}$ at 1.8 days to 12,000 km s$^{-1}$ at 21.4 days. The photospheric temperature evolves from 11,100 K at 1.8 days to 7,200 K at 21.4 days. To fit the high-velocity \ion{H}{i} and \ion{He}{i} lines, the `detach' parameter for these ions was activated, detaching the opacity component of \ion{H}{i} and \ion{He}{i} from the photospheric velocity (v$_\mathrm{phot}$), which defines the inner boundary of the line-forming region. This adjustment allows these lines to form in the outer high-velocity ejecta. The velocities for \ion{H}{i} and \ion{He}{i} from the 1.8 day spectrum are 18,100 km s$^{-1}$ and 16,400 km s$^{-1}$, respectively.

\begin{figure}
	\begin{center}
		\includegraphics[width=\columnwidth]{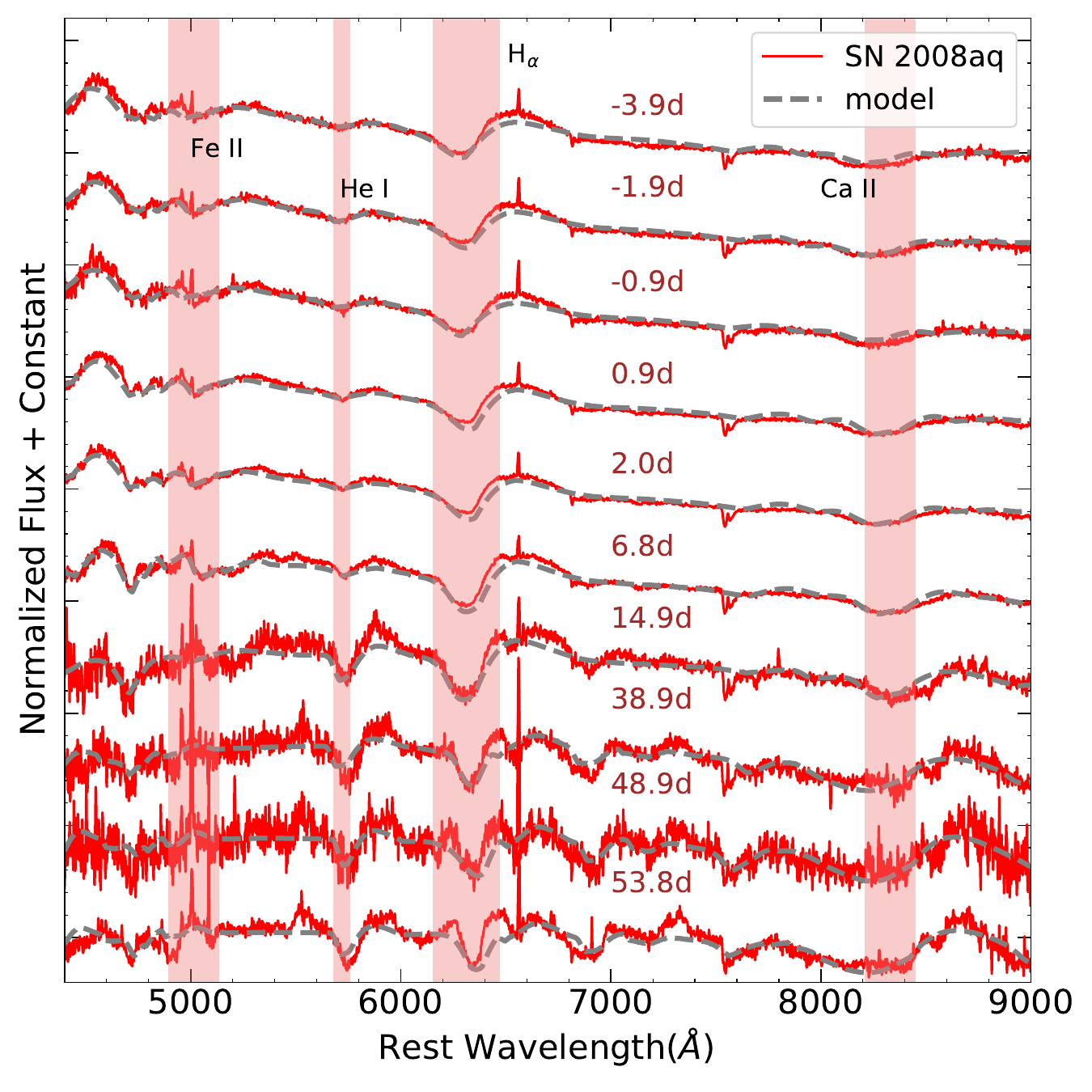}
	\end{center}
	\caption{Spectra of SN~2008aq and their respective models obtained through SYNAPPS are presented in this figure.}
	\label{fig:synapps_2008aq}
\end{figure}

\begin{figure}
	\begin{center}
		\includegraphics[width=\columnwidth]{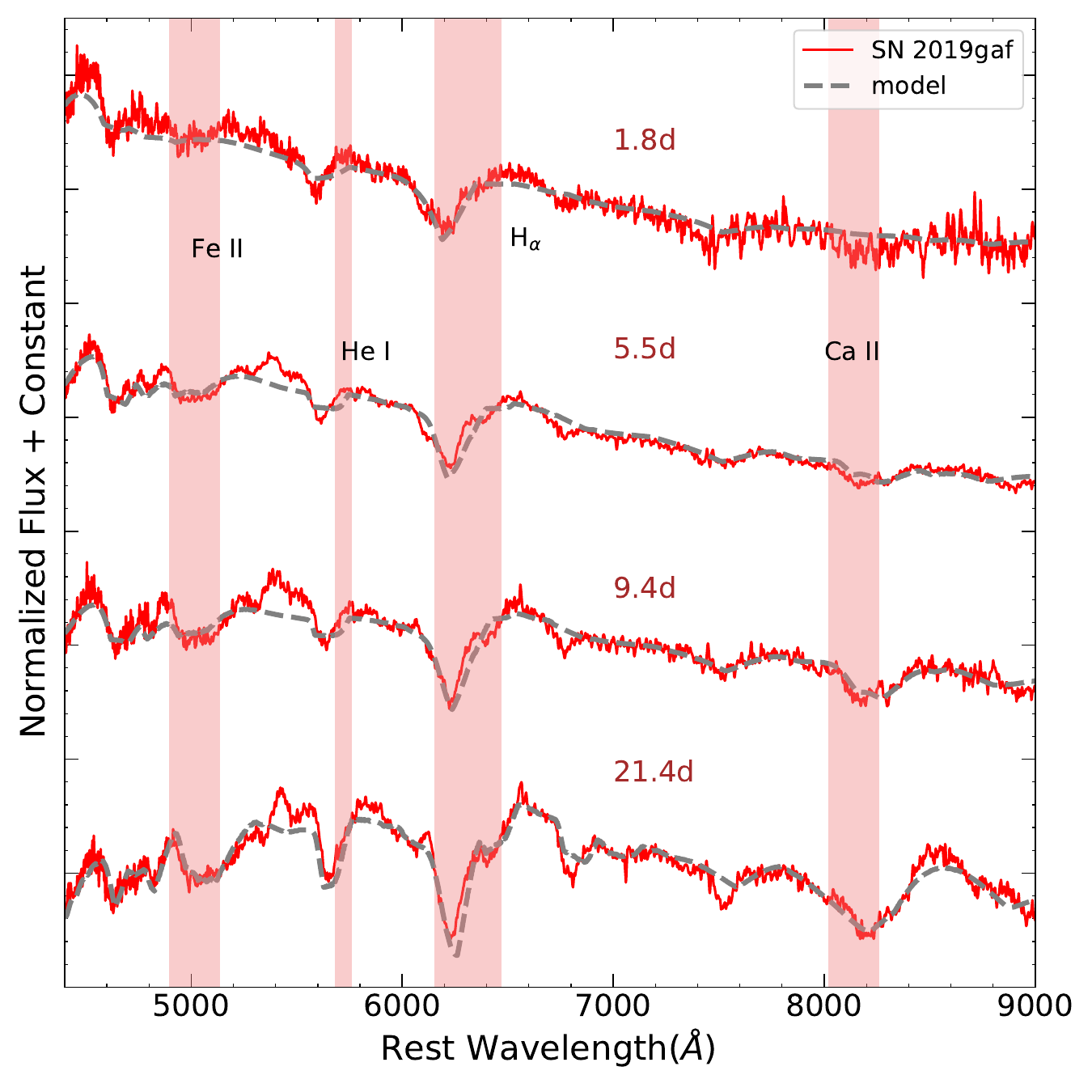}
	\end{center}
	\caption{This figure displays the spectra of SN~2019gaf alongside their corresponding models, as derived using SYNAPPS.}
	\label{fig:synapps_2019gaf}
\end{figure}

\subsection{Expansion velocity}
\label{expansion_velocity}

We measure expansion velocities for SNe~2008aq and 2019gaf using a Gaussian fit to the absorption minima of \ion{H}{$\alpha$} and \ion{He}{i} lines (Figure \ref{fig:expansion_velocity_2008aq_2019gaf}). The velocity evolution of a few other Type IIb SNe is also shown in the same figure. The \ion{H}{$\alpha$} velocity evolution of SN~2008aq is similar to SN~2011dh, slightly higher than SN~2013df, and lower than the rest of the SNe shown in Figure \ref{fig:expansion_velocity_2008aq_2019gaf}. \ion{He}{i} line velocity of SN~2008aq shows similarity with SN~2010as till maximum and then similar to SNe~2011dh and 2013df. Velocity evolution of \ion{H}{$\alpha$} and \ion{He}{i} line for SN~2019gaf is higher than all the Type IIb SNe included in Figure \ref{fig:expansion_velocity_2008aq_2019gaf}. This is consistent with the higher kinetic energy of SN~2019gaf inferred using the light curve modeling (Section \ref{semi_analytical_light_curve_models}). 

\begin{figure}
	\begin{center}
		\includegraphics[width=\columnwidth]{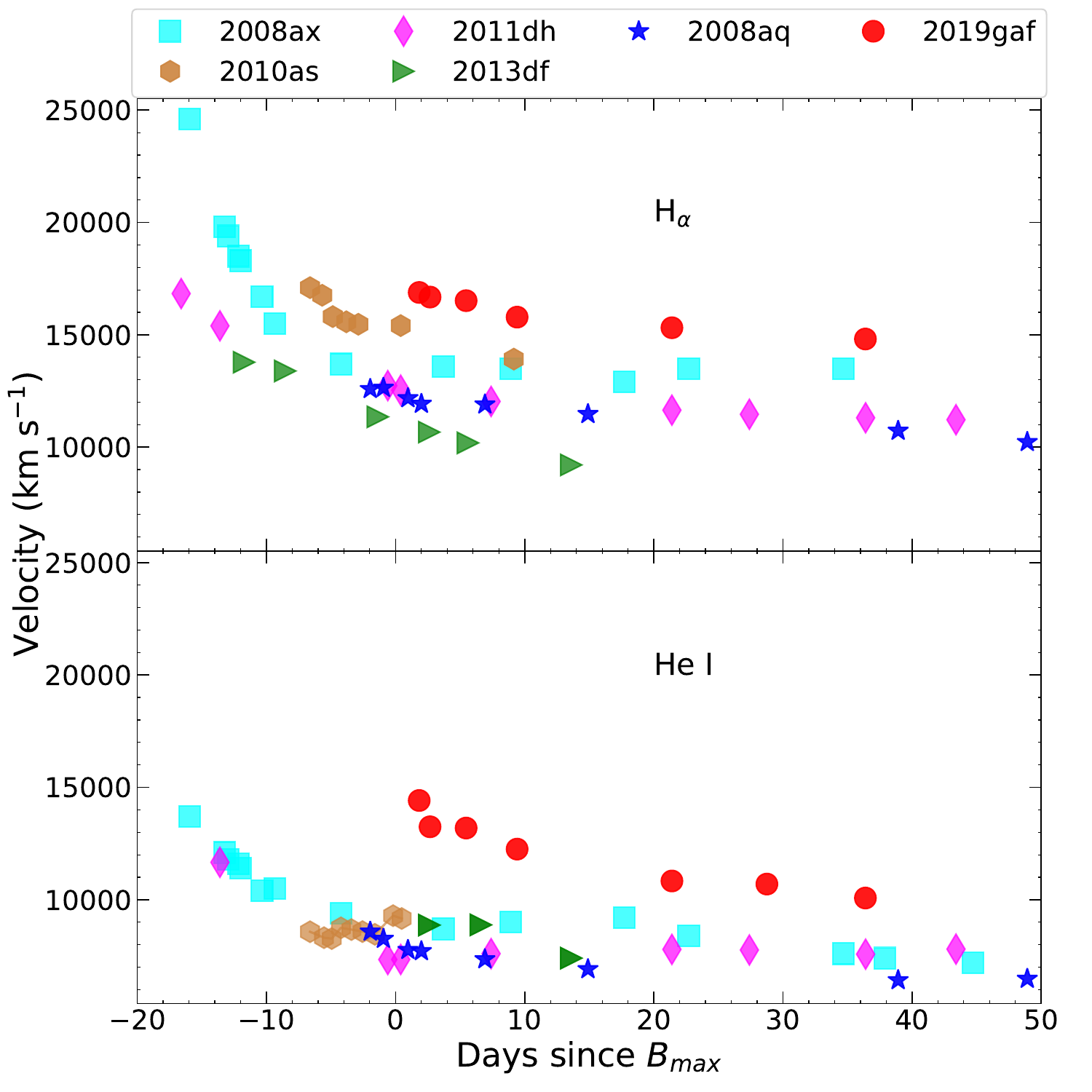}
	\end{center}
	\caption{The figure demonstrates the evolution of photospheric velocities for the \ion{H}{$\alpha$} line and the He I line at 5876 \AA~ for SNe 2008aq and 2019gaf. It also includes velocities estimated for other Type IIb SNe to compare the photospheric velocity evolution of SNe 2008aq and 2019gaf.}
	\label{fig:expansion_velocity_2008aq_2019gaf}
\end{figure}

\section{Explosion parameters and progenitor properties}
\label{explosion_parameters_progenitor_propoerties}

\subsection{Explosion parameters}
\label{explosion_parameters}

\subsubsection{Using Arnett's approximation}
\label{using_arnetts_approximation}

In SESNe, the powering mechanism for the light curves is the energy released from the radioactive decay of $^{56}$Ni to $^{56}$Co to $^{56}$Fe, which means that the peak bolometric luminosity is directly related to the amount of $^{56}$Ni synthesized in the explosion \citep{1982ApJ...253..785A}. To estimate the mass of $^{56}$Ni we have followed expression given by \cite{2005A&A...431..423S} (equation (3) of \citealt{2016MNRAS.458.2973P}). For SNe~2008aq and 2019gaf, the estimated mass of $^{56}$Ni are 0.04 and 0.07 M$_{\odot}$, respectively. To account for the flux contribution from the missing passbands i.e. UV and NIR, \cite{2016MNRAS.458.2973P} quoted 24\% contribution towards the peak luminosity. After adding this, the mass of synthesized $^{56}$Ni for SNe~2008aq and 2019gaf are 0.05 and 0.08 M$_{\odot}$, respectively.     

Using the formulations given by \cite{1982ApJ...253..785A}, assuming spherical symmetry, uniform density, and homologous expansion in the ejecta, we have the following expression for $\tau$$_{m}$ that defines the effective diffusion time scale of the light curve-

\begin{equation}
\tau_{m} = \sqrt(2)(k/(\beta*c))^{0.5}(M_{ej}/v_{ph})^{0.5}
\end{equation}

In this equation, $\kappa$, $\beta$, c, and v$_{ph}$ are the opacity, constant of integration ($\beta$ = 13.8, \citealt{1980ApJ...237..541A,1982ApJ...253..785A}), speed of light, and photospheric velocity at maximum, respectively. In case of SESNe, the opacity $\kappa$ is used as 0.07cm$^{2}$g$^{-1}$ \citep{1992ApJ...394..599C,2000AstL...26..797C,2013MNRAS.434.1098C,2018A&A...609A.136T}. For uniform density and the assumptions mentioned earlier, the constant of integration $\beta$ is estimated as 13.8 by \citet{1980ApJ...237..541A, 1982ApJ...253..785A}.

Also, the kinetic energy using the uniform density of a spherically symmetric ejecta is expressed by

\begin{equation}
    E_k = 0.3M_{ej}*v^{2}_{ph}
\end{equation}

In the case of SNe~2008aq and 2019gaf, $\tau$$_{m}$ are 20 and 17 days, respectively, (assuming the diffusion time scale is equal to rise time), the masses of ejecta are 4.9 and 4.8 M$_{\odot}$, respectively. Here, rise times for both the SNe are calculated using explosion epochs given in Section \ref{distance_extinction_explosion_epoch} and the bolometric maximum. Also, photospheric velocities at maximum for SNe 2008aq and 2019gaf are 11200 and 15000 kms$^{-1}$, respectively.

\subsubsection{Scaling relations:} 

In absence of the detailed hydro-dynamical modelling, the explosion parameters can also be estimated using the scaling relation as shown by \cite{2013MNRAS.432.2463M,2021MNRAS.506.1832M}.  For their objects, the physical parameters were estimated by scaling the explosion parameters of SNe for which these parameters have been estimated using hydrodynamical modelling.  However, it is mentioned also that the rescaling method should be applied with caution, especially if the data coverage is not excellent. It was also emphasized that to have reliable estimates of the explosion parameters, the reference objects should be chosen carefully. There is a similarity in the bolometric light curve of SN 2008aq with SN 2011dh and SN 2019gaf with SN 2010as (Figure \ref{fig:bolometric_plot}). So, we have taken  SNe 20110as \citep{2014ApJ...792....7F}, as the reference SN for SN 2008aq and 2011dh \citep{2012ApJ...757...31B} as reference SN for SN 2019gaf, in our analysis. 

The kinetic energy $E_{k}$ and ejecta mass $M_{\rm ej}$ are determined using the following scaling relations:

\begin{equation}
\frac{E_{\mathrm{k},1}}{E_{\mathrm{k},2}} =
\frac{\tau_{m,1}^{2} \; v_{\mathrm{ph},1}^{3} \; \kappa_{1}^{-1}}
{\tau_{m,2}^{2} \; v_{\mathrm{ph},2}^{3} \; \kappa_{2}^{-1}}
\label{eq:Ek_ratio}
\end{equation}

\begin{equation}
\frac{M_{\mathrm{ej},1}}{M_{\mathrm{ej},2}} =
\frac{\tau_{m,1}^{2} \; v_{\mathrm{ph},1} \; \kappa_{1}^{-2}}
{\tau_{m,2}^{2} \; v_{\mathrm{ph},2} \; \kappa_{2}^{-2}}
\label{eq:Mej_ratio}
\end{equation}

Assuming the optical opacity to be the same for the reference object and our SN, for SN 2008aq, ejecta mass and kinetic energy are estimated as  2.24 M$\odot$  and 1.4$\times$10$^{51}$ erg, respectively. Similarly, for SN 2019gaf these parameters are 6.6 M$\odot$ and 10.8 $\times$10$^{51}$ erg using SN 2010as as reference.

\subsubsection{Semi-analytical light curve models}
\label{semi_analytical_light_curve_models}

We attempt to model the light curve evolution of SNe~2008aq and 2019gaf using a two-component semi-analytical approach presented in \citet{NV5.2016A&A...589A..53N}. This is based on the approach by \citet{NV1.1989ApJ...340..396A}, which has been successively modified in numerous other works (see, for example, \citet{NV2.1993A&A...274..775B, NV3.1993ApJ...414..712P, NV4.2014A&A...571A..77N}). In these models, the ejecta are assumed to be in spherical symmetry with a homologous expansion. In addition, the ejecta is supposed to consist of two independent components, commonly referred to as the inner core part, which has a constant (or flat) density profile and an extended shell defined by an exponential or power law density profile \citep{NV4.2014A&A...571A..77N, NV5.2016A&A...589A..53N}. The incomplete trapping of $\gamma-$rays is incorporated by $\rm A_g$ parameter, which in the overall luminosity could be realized as $\rm L_{bol} = L_{Ni}(1-\exp^{(-Ag/t^2)}) + L_{Rec}$.  We use this semi-analytical approach to estimate the ejecta mass (M$_{ej}$), nickel mass (M$_{Ni}$), explosion energy (E$_{exp}$), and progenitor radius (R$_0$). The errors in the estimated parameters are obtained by varying the best-fitting values to fit the upper and lower limits of errors in the observed luminosities.  With a lack of UV and NIR data, we utilize the blackbody fitted luminosity (L$\rm_{BB}$) obtained using \texttt{SuperBol} as the proxy to the bolometric luminosities (L$_{bol}$) of SN~2008aq and SN~2019gaf. 

\begin{figure}
	\begin{center}
		\includegraphics[width=\columnwidth]{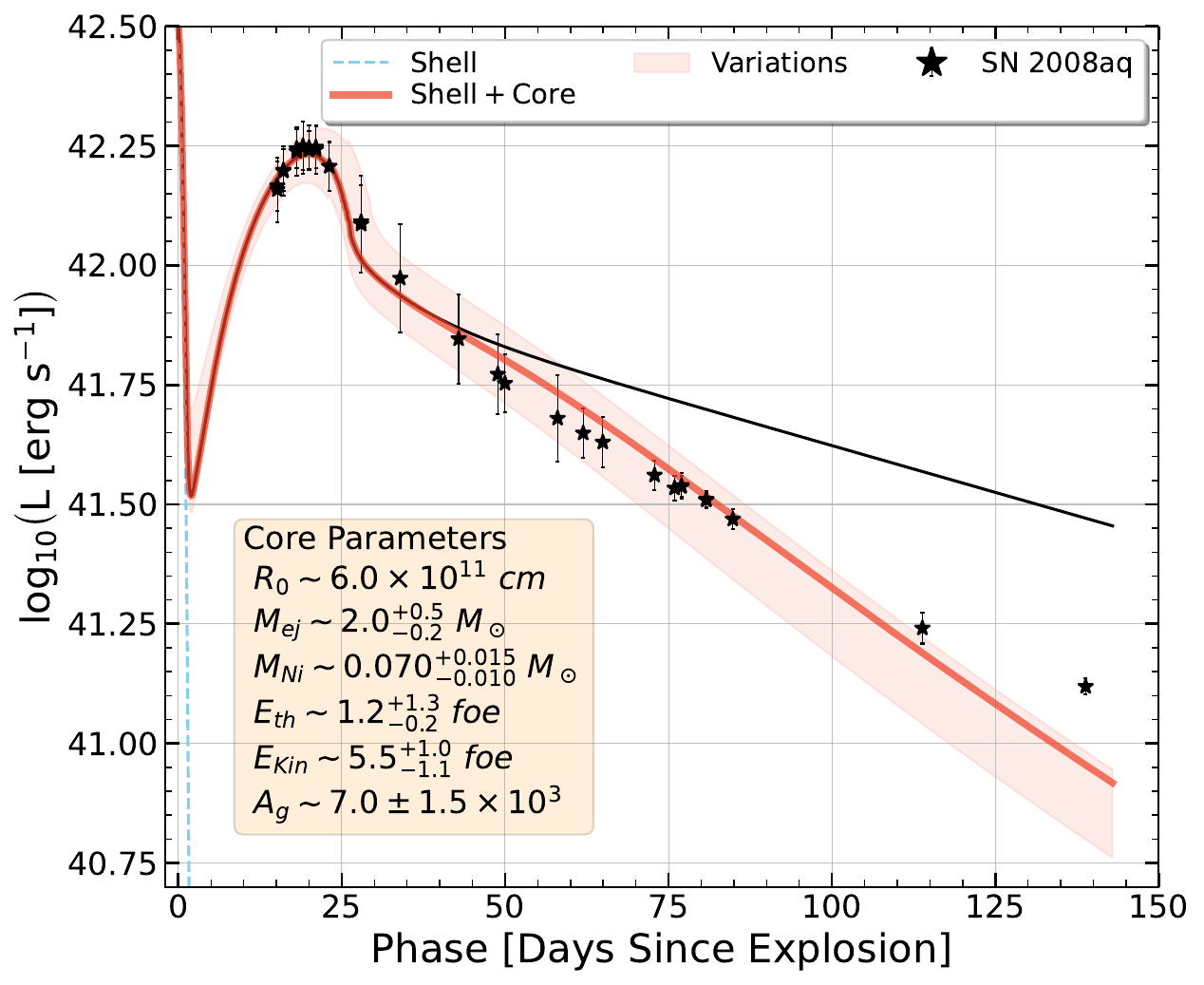}
  \includegraphics[width=\columnwidth]{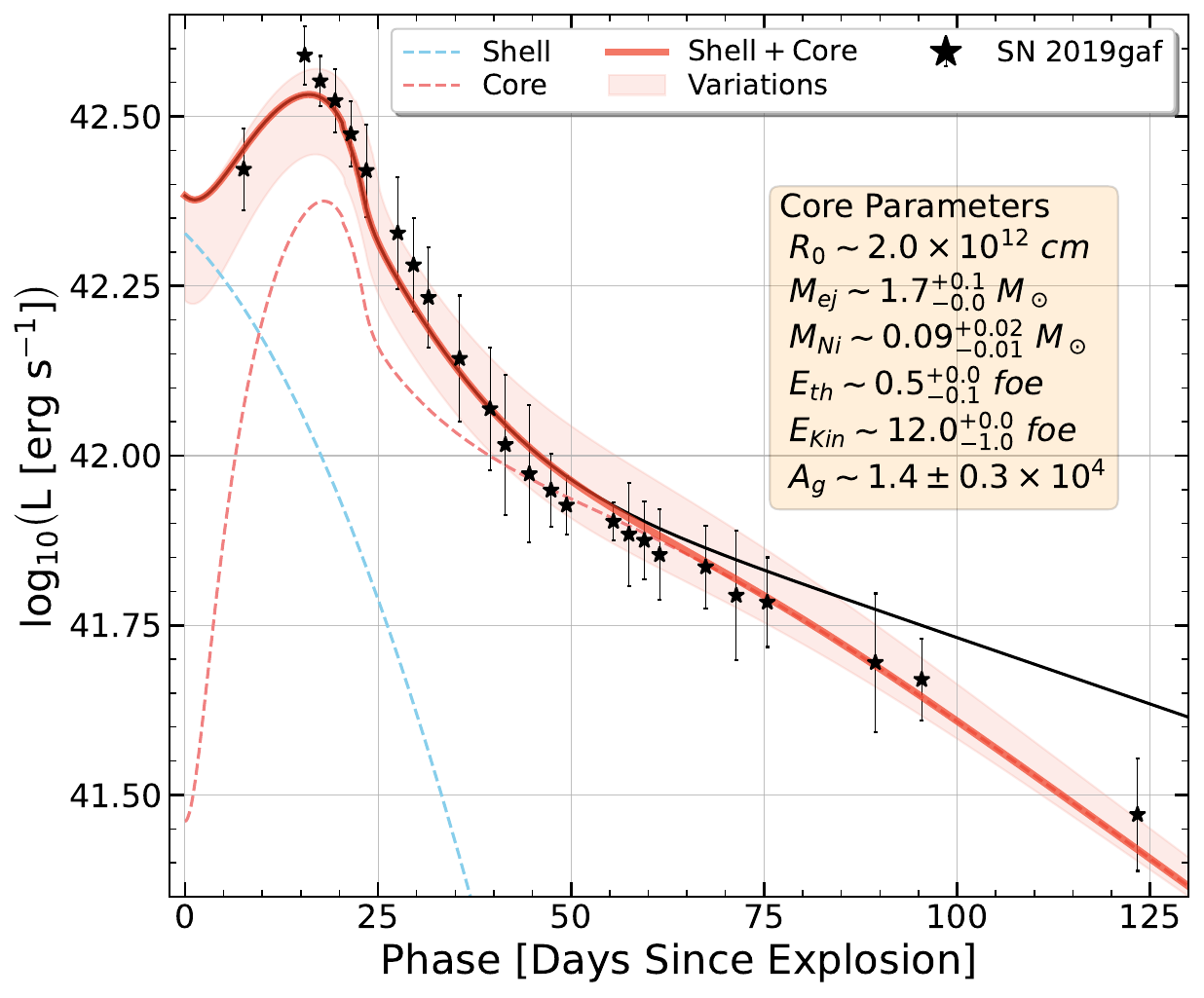}
	\end{center}
	\caption{\textit{Top: }Semi-analytical \citep{NV5.2016A&A...589A..53N} model fit to the bolometric (L$\rm_{BB}$) light curve of SN~2008aq. The solid line represents the model light curve obtained for the best-fitting parameters given in the inset. The shaded region is bound by the variations in the parameters constrained by the errors in the observed light curve. \textit{\textbf{Bottom: }}\textbf{Same as above for SN~2019gaf.} (The black solid line represents the models with complete $\gamma-$ray trapping.)}
	\label{fig:NV2008aq}
\end{figure}

The models that best describe the observed luminosity are shown in Figure~\ref{fig:NV2008aq}, and the core parameters are listed in the inset of the figure. For SN~2008aq, the first observed luminosity epoch is almost 15~days post-explosion, which restricts any constraint on the initial light curve. This part of the light curve is mostly governed by the shell component of the model \citep{NV5.2016A&A...589A..53N}. Due to a lack of early-phase data, we restrict ourselves to core component modeling in the case of SN~2008aq. Hence, we fix the shell component parameters for SN~2008aq so as to not affect the light curve altogether. It implies that the contribution from the shell part is negligible to the total light curve. Hence, we do not vary these parameters during our fitting procedure. For the case of SN~2019gaf, ATLAS photometry has been utilized to obtain the pre-maximum epoch. But again, with insufficient sampling in the pre-maximum epochs, we cannot put any meaningful constraints on the shell part. However, in this case, we do not fix the shell parameters as done in SN~2008aq, owning to the data available around +5~d since the explosion. We vary these as well to obtain the best-fitting modeled light curve. However, to keep the bounds physical, we try to restrict these values close to the parameters obtained in \citet{NV5.2016A&A...589A..53N} for the Type IIb cases.  Eventually, to model both these events, we have mostly focused on the core component, which is also a major flux contributor in the case of stripped progenitors of Type IIb SNe. The best-fitting parameters obtained for both the SNe are presented in Table~\ref{tab:nagyvinko}.

\begin{table}
    \centering
        \caption{Best-fitting parameters obtained using 2-component semi-analytical approach}
        \begin{tabular}{|l|cc|cc|} \hline
        
        {\bf SNe} $\rightarrow$ & \multicolumn{2}{|c|}{\it SN~2008aq} &\multicolumn{2}{|c|}{\it SN~2019gaf}  \\
        \hline
       {\bf Parameters} $\downarrow$  & Core & Shell & Core & Shell \\
         \hline
         R$_0\,[10^{12}~cm]$ &$\sim$0.6 & 12 &$\sim2.0$ & 100 \\
         M$_{ej}\,[M_\odot]$ &$2.0^{+0.5}_{-0.2}$ & 0.001 &$1.7^{+0.1}_{-0.0}$ & 0.5 \\
           M$_{Ni}\,[M_\odot]$ &$0.070^{0.015}_{-0.010}$ & - & $0.09^{+0.02}_{-0.01}$& -  \\
            E$_{Kin}\,[10^{51}~erg]$& $5.5^{+1.0}_{-1.1}$ & 0.0001 &$12.0^{+0.0}_{-1.0}$ & 0.2 \\
             E$_{Th}\,[10^{51}~erg]$& $1.2^{+1.3}_{-0.2}$& 0.001 &$0.5^{+0.0}_{-0.1}$ & 0.04\\
              A$_g\,[10^3~d^2]$& $7.0\pm1.5$& - &$14\pm3$ & - \\
              $k\,[cm^2~g^{-1}]$& 0.2 & 0.3 & 0.3 & 0.3 \\

    \hline
    \end{tabular}
    \label{tab:nagyvinko}
\end{table}

It can be seen that the estimates for M$_{ej}$, arrived at using Arnett's approximation and semi-analytical modeling differ. This could be due to the fact that the Arnett's approximation holds good for a negligible radius (R) and opacity mainly because of electron scattering. Further, the values of $\kappa$ used in Arnett's formulation and the semi-analytical modeling are also different. If we use $\kappa$ as 0.2 and 0.3 cm$^{2}$g$^{-1}$ for SNe 2008aq and 2019gaf (as used in semi analytical modeling prescription), respectively, the ejecta mass estimated using Arnett's formulation are 1.72 and 1.12 M$_{\odot}$. These estimates agree with the values obtained from semi-analytical modeling of the bolometric light curve. Furthermore, the $\rm E_{k}$ and $\rm M_{ej}$ obtained from the scaling relations for SN~2008aq are quite similar to the semi-analytically derived parameters. However, for SN~2019gaf we obtained a consistent explosion energy but with lower ejecta mass when compared with the values obatined from the scaling relations. In the case of scaling relations, we assumed same $\kappa$ for both the reference and observed SNe, however the hydrodynamical models used in the scaling relations are plausibly estimated at different opacities than the average value taken in the semi-analytical formulation. As seen from the Eq~\ref{eq:Ek_ratio} and~\ref{eq:Mej_ratio}, $\rm M_{ej}$ is more sensitive for a given set of opacities. Hence this could be one of the plausible cause of discrepancy in the respective ejecta masses. Additionally, with strong correlation between ejecta mass and opacity there could be a factor of 2 uncertainties in the values estimated from semi-analytical method \citep{NV5.2016A&A...589A..53N}.

\subsection{Progenitor mass using late phase spectra}
\label{Progenitor_mass_using_late_phase_spectra}

Late nebular phase spectral features are an important tool for probing progenitor properties. The ratio of the fluxes of [\ion{O}{i}] and [\ion{Ca}{ii}] lines depends on the ZAMS mass of the progenitor and does not depend upon temperature and density \citep{1989ApJ...343..323F,2004A&A...426..963E}. \cite{2015A&A...579A..95K} presented the ratio of [\ion{O}{i}] and [\ion{Ca}{ii}] lines for core-collapse SNe and marked 1.5 as an arbitrary boundary between the two possible progenitor channels viz., a massive single star progenitor and a less massive star in a binary system. The measured value of [\ion{O}{i}]/[\ion{Ca}{ii}] line ratio in our last observed spectrum (at an epoch of $\sim$ 120 days) of SN~2008aq is 0.82, indicating a binary system as the plausible progenitor channel.
For SN~2019gaf, at an epoch of $\sim$ 112 days, the measured [\ion{O}{i}]/[\ion{Ca}{ii}] ratio is 0.7, indicating that a less massive star in a binary system could be the progenitor of this SN too.

The [\ion{O}{i}] feature is believed to arise from a layer of oxygen formed during the hydrostatic burning phase; hence, the ejected mass of oxygen is directly related to the progenitor's mass. The mass of neutral oxygen can be calculated using the flux of the [\ion{O}{i}] line.  
  
The mass of neutral oxygen M$_{O}$ can be calculated as 

\begin{equation*}
     M_O = 10^8 \times D^2 \times F([\text{O I}]) \times e^{(2.28/T_4)} 
\end{equation*}

where, D is the distance in Mpc, F[O I] is the flux of [\ion{O}{i}] line in the units of erg s$^{-1}$cm$^{-2}$, and T$_{4}$ is the temperature of the [\ion{O}{i}] line emitting region in the unit of 10$^{4}$K and can be calculated using the [\ion{O}{i}] lines fluxes at 5577 and 6300 \AA. In the spectral sequence of SNe~2008aq and 2019gaf, the [\ion{O}{i}] line at 5577 \AA~is not clearly detected, so we have used an upper limit of $\leq$0.1 for the ratio of [\ion{O}{i}] lines at 5577 and 6300~\AA. There are two conditions for this limit, one is low temperature (T$_{4}$ $\leq$ 0.4 K) with high density (N$_{e}$$\geq$10$^{6}$cm$^{-3}$) and another is high temperature (T$_{4}$ = 1.0 K) with low density (N$_{e}$ $\leq$10$^{6}$ cm$^{-3}$). Since low temperature and high densities are consistent with the oxygen region, we have used T$_{4}$ = 0.4K. The measured fluxes of [\ion{O}{i}] line for SNe 2008aq and 2019gaf are 1.93$\times$10$^{-14}$ erg s$^{-1}$cm$^{-2}$ and 1.6$\times$10$^{-14}$ erg s$^{-1}$cm$^{-2}$, respectively.
The calculated mass of neutral oxygen for SNe~2008aq and 2019gaf are 0.65 M$_{\odot}$ and 3.3 M$_{\odot}$, respectively using distance given in Section \ref{distance_extinction_explosion_epoch}. Based on the results of \cite{1996ApJ...460..408T}, ZAMS progenitor masses of 13, 15, 20, and 25 M$_{\odot}$ correspond to oxygen masses of 0.22, 0.43, 1.48, and 3.00 M$_{\odot}$, respectively. Moreover, \cite{1996ApJ...460..408T} presented estimates of helium core masses for progenitor stars with masses of 13, 15, and 25 M$_{\odot}$, yielding values of 3.3, 4.0, and 8.0 M$_{\odot}$, respectively.

The oxygen mass calculated for the SNe suggested that the ZAMS mass of the progenitor for SN~2008aq lies between 15 to 20 M$_{\odot}$, and Helium core mass is between 4 to 8 M$_{\odot}$. For SN~2019gaf, the estimated neutral oxygen mass suggests a ZAMS progenitor mass of $\sim$25 M$_{\odot}$ and a Helium core mass of 8 M$_{\odot}$. The last spectrum of SN~2019gaf available to us has a very poor signal-to-noise ratio. We tried estimating oxygen flux by heavily smoothing this spectrum and hence, the derived oxygen mass and progenitor mass should be taken with caution.

We also compare the late-time spectra of SNe 2008aq and 2019gaf with the models presented in \cite{2015A&A...573A..12J}. \cite{2012A&A...546A..28J} and \cite{2014MNRAS.439.3694J} provided nucleosynthesis yields to constrain the progenitor mass of Type IIP SNe. \cite{2015A&A...573A..12J} expanded the study to include Type IIb SNe using some modifications. These models were constructed for progenitor masses of 12, 13, and 17 M$_{\odot}$ assuming a $^{56}$Ni mass of 0.075 M$_{\odot}$ and distance of 7.8 Mpc. Figures \ref{fig:jerkstrand_2008aq} and \ref{fig:jerkstrand_2019gaf} show the comparison of SNe 2008aq and 2019gaf spectra with the models of \cite{2015A&A...573A..12J}. Our comparison shows that the [\ion{O}{i}] luminosities of SN~2008aq closely match the model for a 13 \(M_\odot\) progenitor. Meanwhile, the [\ion{O}{i}] line luminosities for SN~2019gaf fall between the models for progenitors with masses of 13 and 17 \(M_\odot\). 
Based on the comparison with the models of \cite{2015A&A...573A..12J}, we suggest that the ZAMS mass of the progenitor of SN~2008aq lies between $\sim$ 13 to 20 M$_{\odot}$. For SN~2019gaf, we propose that the ZAMS progenitor mass should range between 13 and 25 \(M_\odot\).

\begin{figure}
	\begin{center}
		\includegraphics[width=\columnwidth]{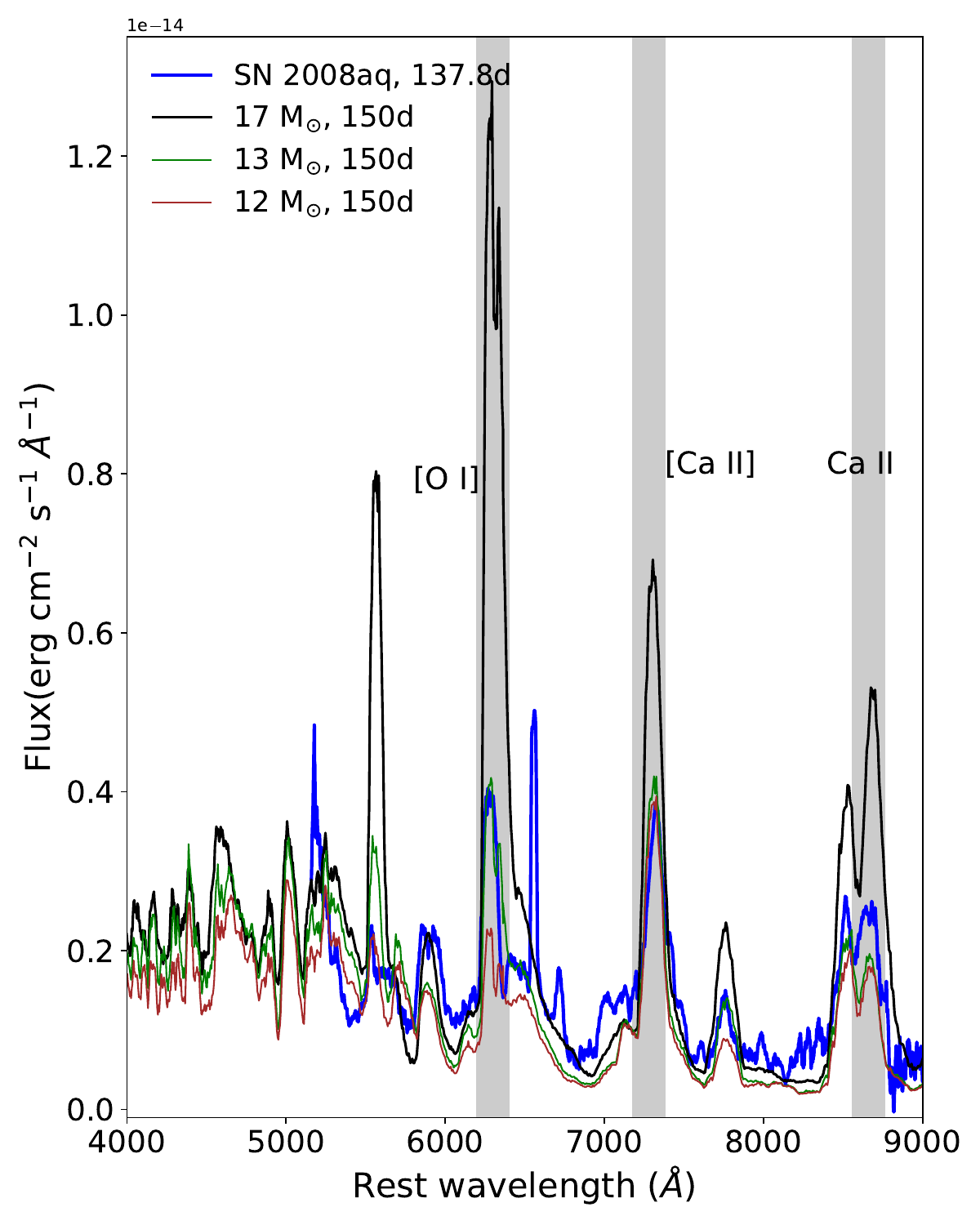}
	\end{center}
	\caption{Comparison of the nebular phase spectrum of SN~2008aq with models for different progenitor mass is shown in this figure.}
	\label{fig:jerkstrand_2008aq}
\end{figure}

\begin{figure}
	\begin{center}
		\includegraphics[width=\columnwidth]{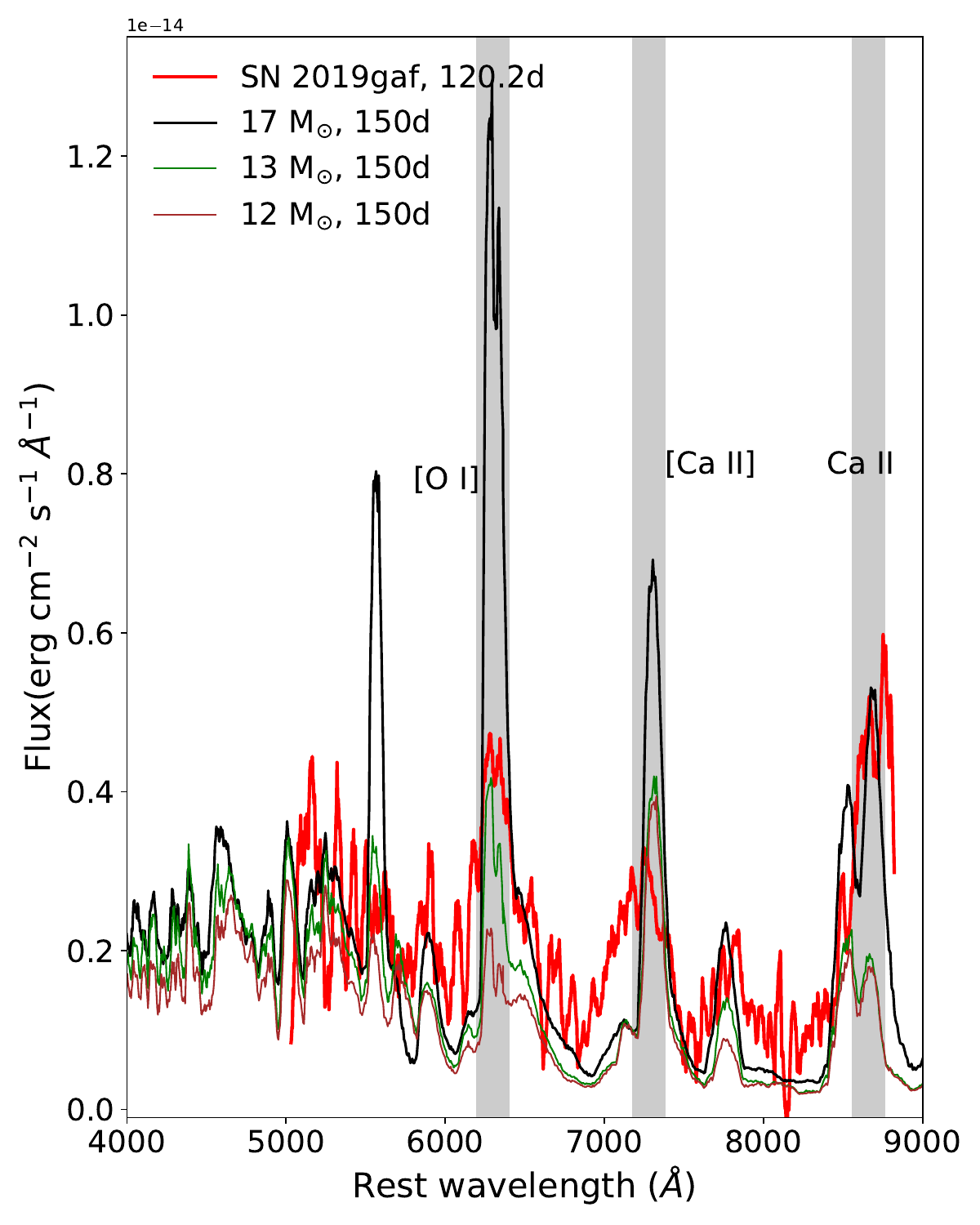}
	\end{center}
	\caption{Same as Figure \ref{fig:jerkstrand_2008aq} but for SN~2019gaf.}
	\label{fig:jerkstrand_2019gaf}
\end{figure}

\section{Summary}
\label{summary}

In this paper, we conducted a detailed analysis of two Type IIb SNe, 2008aq and 2019gaf. SN~2008aq (M$_{B}$ = -16.46$\pm$0.01 mag) belongs to the category of normal Type IIb SNe, while SN~2019gaf (M$_{B}$ = -17.26$\pm$0.08 mag) is situated near the upper luminosity boundary. The B-band light curve decline rate of SN~2008aq shows a faster decline compared to SN~2019gaf. The evolution of colors in both SNe exhibits similarities with other Type IIb SNe used for comparison. The spectral features of both SNe display typical characteristics observed in Type IIb SNe. SN~2019gaf displays a weaker hydrogen feature than typical Type IIb SNe, suggesting that the hydrogen envelope retained by the progenitor before the explosion is thin and extended. Spectral modeling with {\sc SYNAPPS} provides a good fit for the spectral evolution of SNe~2008aq and 2019gaf.

We also performed semi-analytical modeling for both the SNe studied and obtained various explosion parameters. The mass of $^{56}$Ni, ejecta mass, and kinetic energy for SN~2008aq are estimated as  0.07$^{+0.015}_{-0.010}$ M${\odot}$, 2.0$^{+0.5}_{-0.2}$ M${\odot}$ and 5.5$^{+1.0}_{-1.1}$ $\times$ 10$^{51}$ erg, respectively. For SN~2019gaf estimated values of the corresponding parameters are 0.09$^{+0.02}_{-0.01}$ M${\odot}$,  1.7$^{+0.1}_{-0.0}$ M${\odot}$  and 12.0$^{+0.0}_{-1.0}$ $\times$ 10$^{51}$ erg. The $^{56}$Ni and ejecta mass, obtained for both are well within the range obtained for other Type IIb SNe in the literature \citep{2011MNRAS.413.2140T,2013MNRAS.433....2S,2014ApJ...792....7F,2021MNRAS.506.1832M}, whereas the explosion energies are on the higher side especially for the case of SN~2019gaf. However, it has been observed that the semi-analytical modeling approach provides slightly higher energies when compared with other models (e.g., for SN~1993J case \citet{2016MNRAS.457..328L} estimate $\rm \sim1\times10^{51}~erg$ as its $E_{exp}$ whereas \citet{NV5.2016A&A...589A..53N} estimate  $\rm \sim4\times10^{51}~erg$). Another caveat in this modeling is the correlation among the $\kappa, M_{ej},$ and $E_{kin}$ parameters \citep{NV5.2016A&A...589A..53N}, hence only their combination could be well constrained and not individual parameters. We try to vary $\kappa$ and ejecta masses closely to the values estimated in the literature. In addition to this, there is lack of good sampling data especially around the peak for SN~2019gaf which also limit the model's ability to capture true peak and its corresponding phase. Regardless, these parameters could be used as an initial starting point for more detailed hydrodynamical modeling which is beyond the scope of this work.} However, slightly higher energies for SN~2019gaf could be justfied from the expansion velocities of SN~2019gaf which are high, attributing to its high kinetic energy and low ejecta mass as inferred from the semi-analytical modeling. But it is difficult to quantify these at this point.

Based on the nebular phase spectra of both SNe, we have estimated the probable ZAMS progenitor mass to be between $\sim$13 to 20 M$_{\odot}$ for SN~2008aq and between 13 to 25 M$_{\odot}$ for SN~2019gaf. The flux ratio of [\ion{O}{i}] and [\ion{Ca}{ii}] lines, favors a less massive progenitor star in a binary system for both the SNe, however, detailed modeling of the light curve and spectral sequence can further constrain the degree of core stripping, asymmetries in the explosion, and potential contributions from interaction or mixing processes. 
  
\section*{Acknowledgments}

We thank the anonymous referee for the constructive comments and valuable suggestions that have improved the manuscript. MS acknowledges the financial support provided under the National Post Doctoral Fellowship (N-PDF; File Number: PDF/2023/002244) by the Science \& Engineering Research Board (SERB), Anusandhan National Research Foundation (ANRF), Government of India. RD acknowledges funds by ANID grant FONDECYT Postdoctorado Nº 3220449. We acknowledge Wiezmann Interactive Supernova data REPository http://wiserep.weizmann.ac.il (WISeREP) \citep{2012PASP..124..668Y}. This research has made use of the CfA Supernova Archive, which is funded in part by the National Science Foundation through grant AST 0907903. This research has made use of the NASA/IPAC Extragalactic Database (NED) which is operated by the Jet Propulsion Laboratory, California Institute of Technology, under contract with the National Aeronautics and Space Administration. This work makes use of data obtained with the LCO Network. We thank the staff of IAO, Hanle, and CREST, Hosakote, that made these observations possible. The facilities at IAO and CREST are operated by the Indian Institute of Astrophysics, Bangalore. The LCO group were supported by NSF Grants AST-1911151 and AST-1911225. 

\section*{ Data availability} The photometric and spectroscopic data presented in this paper will be available upon request from the corresponding author.
    











\bibliographystyle{mnras}
\bibliography{ms}

\bsp	
\label{lastpage}
\end{document}